\documentclass[a4paper,11pt]{article}
\pdfoutput=1 

\usepackage{jheppub} 

\usepackage[T1]{fontenc} 
\usepackage{graphicx}
\usepackage{MnSymbol}
\graphicspath{{./figures/}}

\newcommand{\be}{\begin{eqnarray}}
\newcommand{\ee}{\end{eqnarray}}
\newcommand{\bdm}{\begin{displaymath}}
\newcommand{\edm}{\end{displaymath}}
\newcommand{\ds}{\displaystyle}
\newcommand{\ba}{\begin{array}}
\newcommand{\ea}{\end{array}}
\newcommand{\pa}[1]{\left(#1\right)}

\newcommand{\dpa}{\partial}

\newcommand{\vep}{\varepsilon}
\newcommand{\z}[1]{\zeta_{#1}}
\allowdisplaybreaks

\title{\boldmath Effective field theory approach to the gravitational
  two-body dynamics, at fourth post-Newtonian order and quintic in the Newton constant}

\author[a]{Stefano Foffa}
\author[b]{Pierpaolo Mastrolia}
\author[c]{Riccardo Sturani}
\author[d]{Christian Sturm}

\affiliation[a]{D\'epartement de Physique Th\'eorique and Centre for Astroparticle Physics, Universit\'e de 
             Gen\`eve, CH-1211 Geneva, Switzerland}
\affiliation[b]{Dipartimento di Fisica ed Astronomia, Universit\`a di
  Padova, Via Marzolo 8, 35131 Padova, Italy\\
INFN, Sezione di Padova, Via Marzolo 8, 35131 Padova, Italy}
\affiliation[c]{International Institute of Physics (IIP),
Universidade Federal do Rio Grande do Norte (UFRN) CP 1613, 59078-970 Natal-RN Brazil}
\affiliation[d]{{Universit{\"a}t W{\"u}rzburg, Institut f{\"u}r Theoretische Physik und Astrophysik, Emil-Hilb-Weg 22, D-97074 W{\"u}rzburg, Germany}}

\emailAdd{stefano.foffa@unige.ch}
\emailAdd{pierpaolo.mastrolia@pd.infn.it}
\emailAdd{riccardo@iip.ufrn.br}
\emailAdd{Christian.Sturm@physik.uni-wuerzburg.de}

\abstract{Working within the post-Newtonian (PN) approximation to General Relativity,
we use the effective field theory (EFT) framework to study the conservative dynamics
of the two-body motion at fourth PN order, at fifth order in the Newton constant.
This is one of the missing pieces preventing the computation of the
full Lagrangian at fourth PN order using EFT methods.
We exploit the analogy between diagrams in the EFT gravitational
theory and 2-point functions in massless gauge theory, to address the
calculation of 4-loop amplitudes by means of standard multi-loop
diagrammatic techniques.
For those terms which can be directly compared, our result confirms the findings of previous 
studies, performed using different methods.
}

\begin{document} 
\maketitle
\flushbottom

\section{Introduction}
\label{sec:intro}
The post-Newtonian (PN) approximation to the 2-body problem in General
Relativity has been subject of intense investigation in the last decades
as it describes the dynamics of gravitationally bound binary systems 
in the weak curvature, slow velocity regime, reviewed in 
\cite{lrr-2002-3,Futamase:2007zz} and \cite{Porto:2016pyg}.

From the phenomenological point of view its results have been of paramount
importance in constructing the waveforms which have been
eventually used as templates \cite{Taracchini:2012ig,Schmidt:2014iyl}
for the LIGO/Virgo data analysis pipeline leading to the detection
\cite{Abbott:2016blz},
along with numerical simulation allowing to solve for the space time in the
strong curvature regime \cite{PhysRevLett.111.241104} and earlier in the
analysis of the Hulse-Taylor pulsar arrival times \cite{Taylor:1982zz,Damour:1983tz}.

Interferometric detectors of gravitational waves are particularly sensitive
to the time varying phase of the signal of coalescing binaries, which thus
must be computed with better than ${\cal O}(1)$ precision \cite{Cutler:1992tc}.
Such a phase can be determined from short-circuiting the information of the
energy and luminosity function of binary inspirals with at least 3PN order
accuracy.

Focusing on the conservative sector of the two body problem without
spins (see \cite{Porto:2016pyg} for results involving spins),
we recall that within the EFT formalism, initially proposed in \cite{Goldberger:2004jt}
and reviewed in \cite{Goldberger:2007hy,Foffa:2013qca,Rothstein:2014sra,Porto:2016pyg},
the 1PN, 2PN \cite{Gilmore:2008gq} and 3PN \cite{Foffa:2011ub} dynamics have
been computed, reproducing results obtained with more
traditional methods; moreover the 4PN Lagrangian, quadratic in the
Newton constant $G_N$, was first derived in the EFT framework
\cite{Foffa:2012rn}.

The complete 4PN dynamics has been obtained recently by two groups 
within the Arnowitt-Deser-Misner Hamiltonian formalism \cite{Damour:2014jta,Damour:2015isa}
and by iterating the PN equation in the harmonic gauge in 
\cite{Bernard:2015njp,Bernard:2016wrg}; in both approaches an arbitrary coefficient has
been fixed by using results for the gravitational wave tail effect from
self-force computations \cite{Blanchet:2010zd,LeTiec:2011ab,Bini:2013zaa}.
It is worth mentioning that the two results did not initially agree at orders $G_N^4$ and $G_N^5$ and, as it is argued in 
\cite{Damour:2016abl}, the discrepancy has been
overcome by a suitable regularization of the infrared and ultraviolet
divergencies in the approach based on the equations of motion, although the new regularization could not fix yet the value of the
second ambiguity parameter in \cite{Bernard:2016wrg}.

This work goes in the direction of providing a third-party computation with an independent methodology by filling one of the missing pieces to obtain the full 4PN result within EFT methods.
Using the virial relation $v^2\sim G_NM/r$,
being $r$ and $v$ respectively the relative distance and velocity
of the binary 
constituents with $M$ the total mass, the terms contributing to the
4PN order dynamics can be parametrized as $G_N^{5-n}v^{2n}$ with $0\leq n\leq 5$,
the leading term being the Newtonian potential, scaling simply as $G_N$.
By following on the way paved by \cite{Foffa:2012rn}, 
we present in this work some results concerning the $G_N^5$ order.

The Lagrangian contains in general terms with
high derivative of the dynamical variables: it is however possible to keep the
equations of motion of second order without altering the dynamics by
adding to the Lagrangian terms \emph{quadratic} at least in the equations of motions
tuned to cancel the high derivative terms at the price of introducing additional
terms with higher $G_N$ powers, according to the standard procedure first
proposed in \cite{Damour:1985mt} and dubbed \emph{double zero} technique.
The $G_N^5$ sector of the Lagrangian receives contributions from $G_N$, $G_N^2$
and $G_N^3$ Lagrangian terms which are at least quadratic in accelerations
(computed in \cite{Foffa:2012rn} up to $G_N^2$) via the \emph{double zero} trick, as well as from \emph{genuine}
$G_N^5$ terms:
in the present article, we focus on the genuine $G_N^5$ contribution, that is terms that
do not contain {\em  ab initio} any power of velocity $v$ or
acceleration ${\dot v}$, 
and leave the very last contribution, coming  from 
${\cal O}( G_N^3\dot{v}^2)$ terms, to a forthcoming paper
dedicated to the whole $G_N^3$ sector.

In this work, we evaluate the 50 diagrams contributing to the classical
effective Lagrangian in the gravitational theory at order $G_N^5$. 
They are non-trivial integrals over 3-momenta which can be computed 
by means of multi-loop diagrammatic techniques.
We exploit the analogy between diagrams in the EFT gravitational
theory and diagrams corresponding to 2-point functions in massless gauge theory, 
to address the calculation of the ${\cal O}( G_N^5)$ diagrams as
2-point 4-loop dimensionally regulated integrals in $d$ dimensions. In particular, we use integration-by-parts identities (IBPs)
\cite{Tkachov:1981wb,Chetyrkin:1981qh,Laporta:2001dd} in two ways: 
according to the topology of the graph, IBPs allow to carry out the multiloop integration recursively
loop-by-loop; alternatively, they can be used to express the result of
the amplitudes as linear combination of irreducible integrals, known
as {\it master integrals} (MIs). 
The latter are evaluated independently.
The contribution to the three-dimensional Lagrangian coming from each graph is then 
determined by taking the $d \to 3$ limit of the Fourier transform to position-space.


The paper is organized as follows. In sec.~\ref{sec:method} we review the EFT
formalism applied to the two-body dynamics in the PN approximation to General
Relativity and in sec.~\ref{sec:results} we present the details
of the 4PN computation at $G_N^5$ order.
We summarize in sec.~\ref{sec:discussion} and conclude in sec.~\ref{sec:conclusion}. Appendix~\ref{app:masters} contains
the expressions of the master integrals needed for the computation,
in Appendix~\ref{app:50amplitudes} we give the contribution to the Lagrangian coming
from the individual diagrams and in Appendix~\ref{app:a33_a50} details
of the computation of selected amplitudes are reported.

\section{The method}
\label{sec:method}
The application of the EFT framework to post-Newtonian calculations in binary dynamics has
now been extensively investigated.
It was first formulated in this context in \cite{Goldberger:2004jt} and subsequently applied to various aspects of the binary problem
(see reviews \cite{Foffa:2013qca, Porto:2016pyg} and references therein).

We summarize here the basic features of this approach, along the lines and notations of \cite{Foffa:2011ub, Foffa:2012rn}, while referring the reader to the literature for a more complete account. The starting point is the action
\be
\label{action}
S = S_{bulk}+S_{pp}\,,
\ee
with the world-line point particle action representing the binary components
(we only consider here spinless point masses and neglect tidal effects)
 \be
\label{az_pp}
S_{pp}=-\sum_{i=1,2} m_i\int {\rm d}\tau_i = -\sum_{i=1,2} m_i\int \sqrt{-g_{\mu\nu}(x_i) {\rm d}x_i^\mu {\rm d}x_i^\nu}\,,
\ee
as well as the usual Einstein-Hilbert action\footnote{
We adopt the ``mostly plus'' convention
$\eta_{\mu\nu}\equiv {\rm diag}(-,+,+,+)$, and the Riemann and Ricci tensors are
defined as $R^\mu_{\nu\rho\sigma}=\dpa_\rho\Gamma^\mu_{\nu\sigma}+
\Gamma^\mu_{\alpha\rho}\Gamma^\alpha_{\nu\sigma}-\rho\leftrightarrow\sigma$, 
$R_{\mu\nu}\equiv R^\alpha_{\mu\alpha\nu}$. }
plus a gauge fixing term
\be\label{az_bulk}
S_{bulk}= 2 \Lambda^2 \int {\rm d}^{d+1}x\sqrt{-g}\left[ R(g)-\frac{1}{2}\Gamma_\mu\Gamma^\mu\right]\,,
\ee
which corresponds to the same harmonic gauge condition adopted in
refs.~\cite{lrr-2002-3,Bernard:2015njp},
where $\Gamma^\mu \equiv g^{\rho\sigma}\Gamma^\mu_{\rho\sigma}$.
Here $\Lambda^{-2}\equiv 32 \pi G_N L^{d-3}$, with $G_N$ the 3-dimensional Newton constant and $L$ an arbitrary length scale which  keeps the correct dimensions of $\Lambda$ in dimensional regularization, and always cancels out in the expression of physical observables.\\
In this framework, a Kaluza-Klein (KK) 
parametrization  of the metric \cite{Kol:2007bc,Kol:2007rx} is usually adopted (a somehow similar 
parametrization was first applied within the framework of a PN calculation in
\cite{Blanchet:1989ki}):
\be
\label{met_nr}
g_{\mu\nu}=e^{2\phi/\Lambda}\pa{
\ba{cc}
-1 & A_j/\Lambda \\
A_i/\Lambda &\quad e^{- c_d\phi/\Lambda}\gamma_{ij}-
A_iA_j/\Lambda^2\\
\ea
}\,,
\ee
with, $\gamma_{ij} \equiv \delta_{ij}+\sigma_{ij}/\Lambda$,
$c_d \equiv 2\frac{(d-1)}{(d-2)}$ and $i,j$ running over the $d$ spatial dimensions.
The field $A_i$ is not actually needed in the present computation, so it will henceforth be set to zero;
we refer to \cite{Foffa:2011ub} for the general treatment and formulae including $A_i$.

In terms of the metric parametrization (\ref{met_nr}), with $A_i=0$,
each world-line coupling to the gravitational degrees of freedom
$\phi$, $\sigma_{ij}$  reads
\renewcommand{\arraystretch}{1.4}
\be
\label{matter_grav}
S_{pp}=-m\ds \int {\rm d}\tau = \ds-m\int {\rm d}t\ e^{\phi/\Lambda}
\sqrt{1
-e^{-c_d \phi/\Lambda}\pa{v^2+\frac{\sigma_{ij}}{\Lambda}v^iv^j}}\,,
\ee
\renewcommand{\arraystretch}{1.4}
and its Taylor expansion provides the various particle-gravity vertices of the EFT.

Also the pure gravity sector $S_{bulk}=S_{EH}+ S_{GF}$ can be explicitly written
in terms of the KK variables; we report here only those terms which are needed
for the present calculation\footnote{It is understood that spatial indices in
this expression, including those implicit in terms carrying a $(\vec\nabla)^2$,
are contracted by means of the spatial metric $\gamma_{ij}$, which implies the
appearance of extra $\sigma$ fields, e.g.
$(\vec{\nabla}\sigma)^2\equiv\gamma^{ab}\gamma^{cd}\gamma^{ij}\sigma_{ab,i}\sigma_{cd,j}$ and $\gamma^{ij}=(\gamma^{-1})_{ij}$ (and on the second line $\sigma^{ij}=\sigma_{ij}$, $\sigma=\delta^{ij}\sigma_{ij})$.}:
\renewcommand{\arraystretch}{1.4}
\be
\label{bulk_action}
\ds S_{bulk} &\supset& \ds \int {\rm d}^{d+1}x\sqrt{\gamma}
\left\{\frac{1}{4}\left[(\vec{\nabla}\sigma)^2-2(\vec{\nabla}\sigma_{ij})^2\right]- c_d (\vec{\nabla}\phi)^2\right.\nonumber\\
&-&\left.\frac{1}{\Lambda}\left(\frac{\sigma}{2}\delta^{ij}-\sigma^{ij}\right)
\left({\sigma_{ik}}^{,l}{\sigma_{jl}}^{,k}-{\sigma_{ik}}^{,k}{\sigma_{jl}}^{,l}+\sigma_{,i}{\sigma_{jk}}^{,k}-\sigma_{ik,j}\sigma^{,k}
\right)\right\}\,.
\ee
\renewcommand{\arraystretch}{1.}

\begin{figure}[h]
\begin{center}
\includegraphics[width=1.0\linewidth]{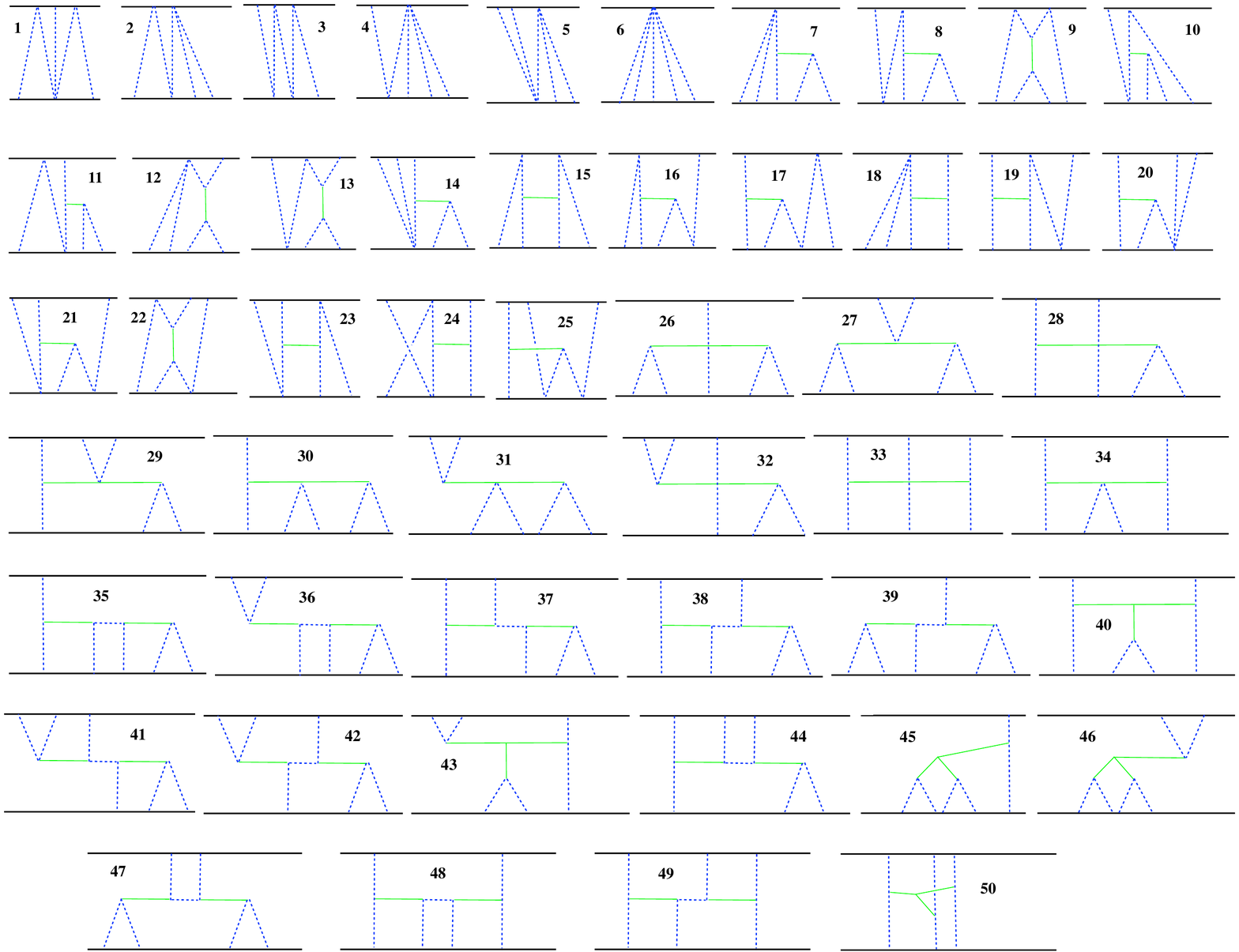}
\caption{The diagrams contributing at order $G^5_N$. As in the EFT
  approach the massive objects are non-dynamical, the horizontal black
  lines have to be seen as classical sources, and not as
  propagators. Green solid lines stand for $\sigma$ field propagators,
  blue dashed lines for $\phi$ fields.}
\label{diaG5}
\end{center}
\end{figure}

\label{method}
The 2-body effective action can be found by integrating out the gravity fields from the above-derived actions
\begin{equation}
\label{eq:seff}
 \exp[\text{i} S_{eff}]=\int D\phi D\sigma_{ij} \exp[\text{i}(S_{bulk}+ S_{pp})]\,.
\end{equation}
As usual in field theory, the functional integration can be perturbatively 
expanded in terms of Feynman diagrams involving the gravitational 
degrees of freedom as internal lines only
\footnote{As we focus on the conservative part of the dynamics, we are not
interested in diagrams where gravitational radiation is released to infinity,
even though \emph{tail} effects \cite{Blanchet:1987wq} involving emitted and absorbed radiation
are relevant at $G_N^2$ order also in the conservative sector.}, 
regarded as dynamical fields emitted and absorbed by the point particles which 
are taken as non-dynamical sources.

In order to make manifest the $v$ scaling necessary to classify the results according to the PN hierarchy, it is convenient to work with the 
space-Fourier transformed fields
\renewcommand{\arraystretch}{1.4}
\be
\label{Fourk}
W^a_p(t)  \equiv \ds\int {\rm d}^dx\, W^a(t,x) e^{-\text{i} p\cdot x}\,\quad
 {\rm with\ } W^a=\{\phi,\sigma_{ij}\}\,.
\ee
\renewcommand{\arraystretch}{1}
The fields defined above are the fundamental variables in terms of
which we are going to construct the Feynman graphs; the action governing
their dynamics can be found from eqs.~(\ref{matter_grav},\ref{bulk_action}).

The next step is to lay down all the diagrams which contribute at this ${\cal O}(G_N^5)$ in the static limit,
following the rule that each vertex involving $n$ gravitational fields carries a factor $G_N^{n/2-1}$ if it is a bulk one,
and a factor $G_N^{n/2}$ if it is attached to an external particle.

The diagrams in fig.~\ref{diaG5} schematically represent the exchange
of gravitational potential modes through the field $\phi$ (blue dotted lines)
and $\sigma_{ij}$ (green solid line) which mediate the
gravitational interaction. Massive objects represented by the thick horizontal
black solid line are non-dynamical sources or sinks of gravitational modes.
Their dynamics is described by the world line $S_{pp}$ hence no massive
particle propagator is present in between two different insertions of 
gravitational modes on the same particle.

The amplitudes corresponding to each diagram can be built from the
Feynman rules in momentum-space derived from ${\cal S}_{pp}$, ${\cal S}_{bulk}$.
By looking in particular at the quadratic parts, one can explicitly write the 
propagators:
\be
\label{propagators}
\ds P[W^a_p(t_a)W^b_{p'}(t_b)]&=&\ds \frac{1}{2} P^{aa}\delta_{ab}
\ds (2\pi)^d\delta^{d}(p+p'){\cal P}(p^2,t_a,t_b)\delta(t_a-t_b)\,,
\ee
where $P^{\phi\phi}=-\frac{1}{c_d}$, 
$P^{\sigma_{ij}\sigma_{kl}}=-\left(\delta_{ik}\delta_{jl}+\delta_{il}\delta_{jk}+(2-c_d)\delta_{ij}\delta_{kl}\right)$ 
and
\be\label{timeprop}
{\cal P}(p^2,t_a,t_b)=\frac{\text{i}}{p^2-\partial_{t_a}\partial_{t_b}}\simeq
\frac{\text{i}}{p^2}\ee
has been truncated to its instantaneous non-relativistic part.
The terms involving time derivatives (which acting on the 
$e^{ip\cdot x}$, generate extra factors of $v$) can be
indeed neglected. In fact, in the present work, we are interested in the
pure 4PN $G_N^5$ contribution, which, by power counting, can be accessed in the limit of zero
velocity and instantaneous interactions. 
In other words, gravitational mode momenta
have scaling of the types $(v/r,1/r)$, therefore the temporal
component of their momenta can be neglected, since we are computing the $G_N^5 v^0$ sector.

From the previous discussion, one can derive the following Feynman rules, respectively for the $\phi$-propagator,
\begin{equation}
\ba{ccl}
\begin{minipage}{3.0cm}
\begin{center}
\includegraphics[width=3.0cm]{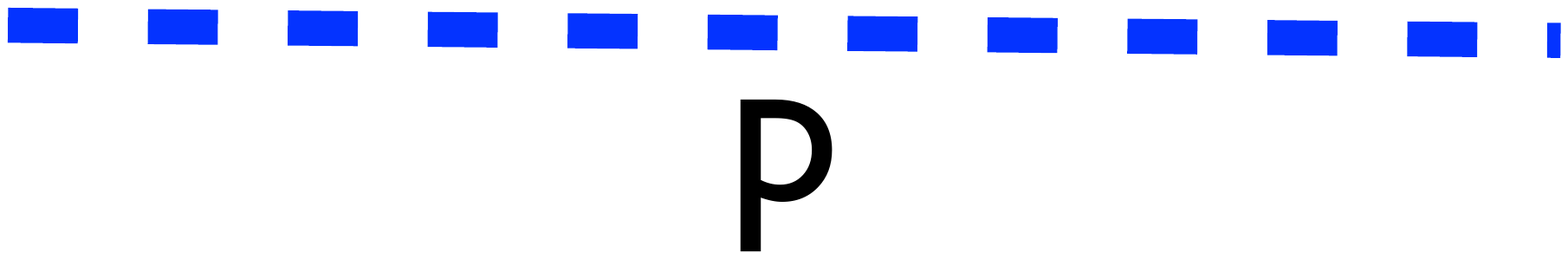}
\end{center}
\end{minipage}
&\rightarrow&\displaystyle - \frac{\text{i}}{2 c_d p^2}
\ea
\end{equation} 
and for the $\sigma$-propagator,
\begin{equation}
\ba{ccl}
\begin{minipage}{3.0cm}
\begin{center}
\includegraphics[width=3.0cm]{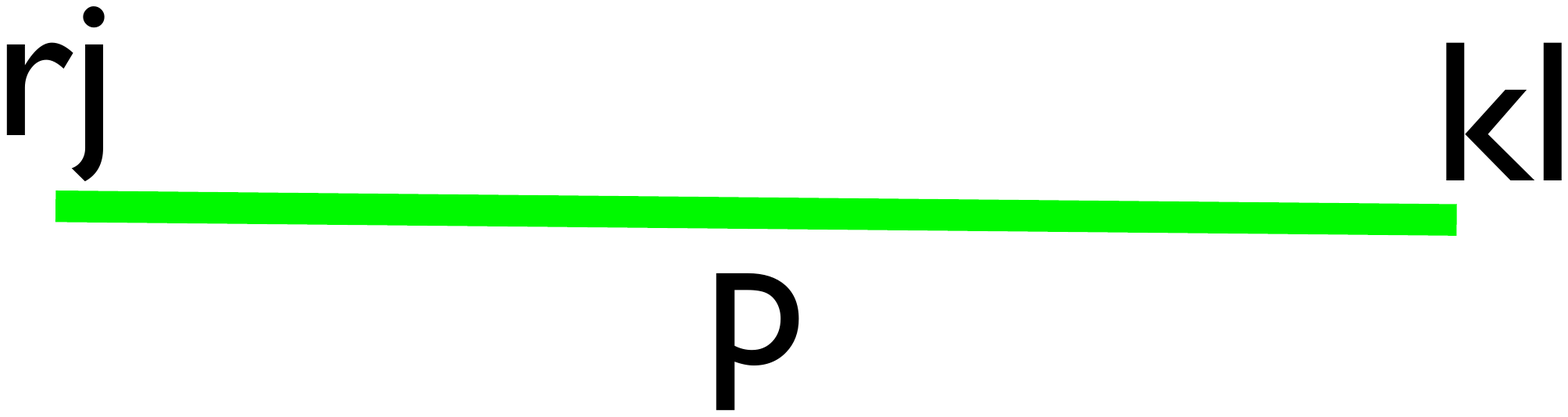}
\end{center}
\end{minipage}
&\rightarrow&\displaystyle \frac{\text{i}P^{\sigma_{rj}\sigma_{kl}}}{2 p^2}\,.
\ea
\end{equation}
The Feynman rules for the interaction vertices can be derived in a similar fashion and are reported below:
\begin{eqnarray}
\begin{minipage}{2.0cm}
\begin{center}
\includegraphics[width=2.0cm]{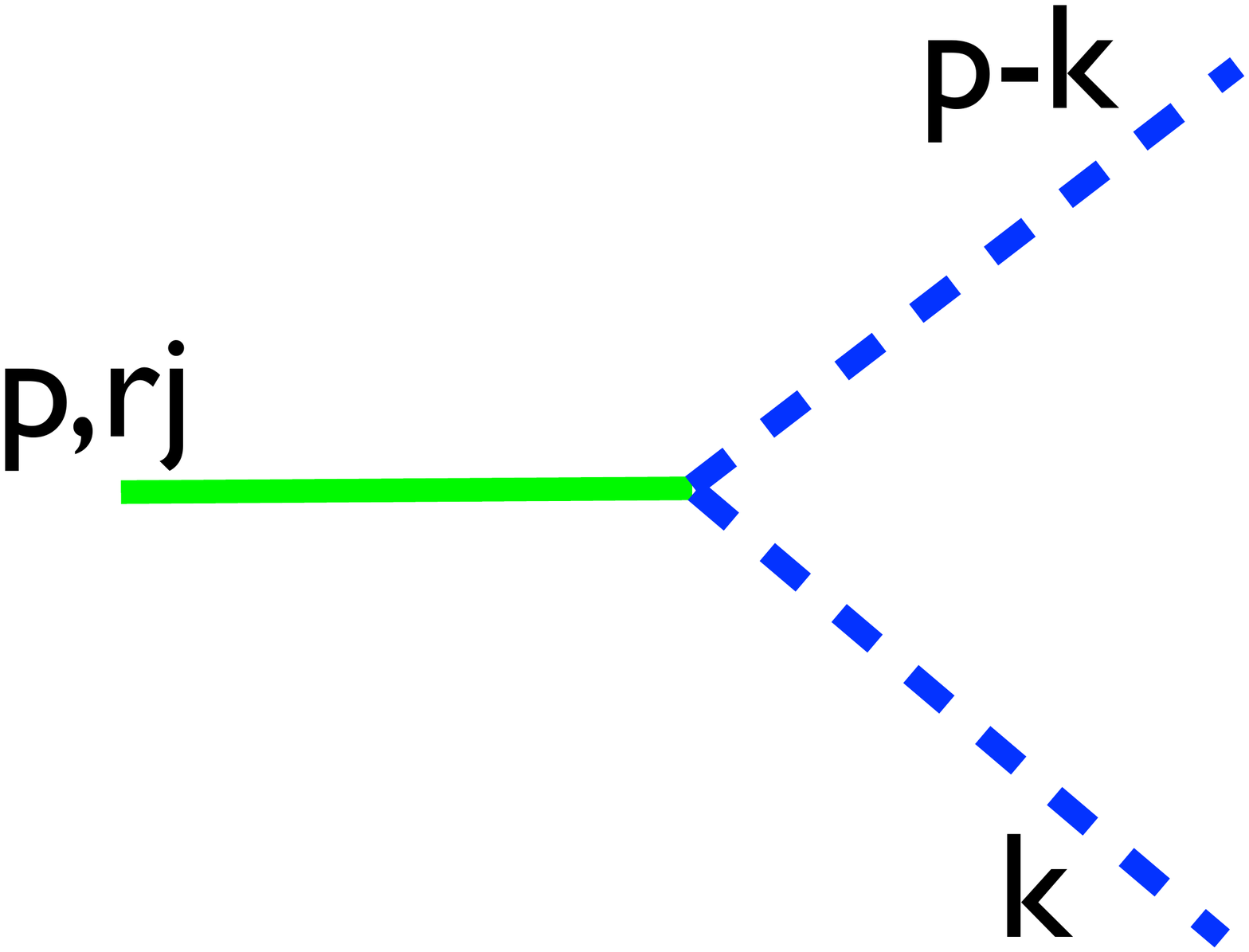}
\end{center}
\end{minipage}
&\rightarrow&\text{i} \frac{2c_d}{\Lambda} \left[\frac{1}{2}(p-k) \cdot
  k\delta^{rj}-k^r (p-k)^j+\Big(r\leftrightarrow j\Big)\right],\nonumber
\\
\begin{minipage}{2.0cm}
\begin{center}
\includegraphics[width=2.0cm]{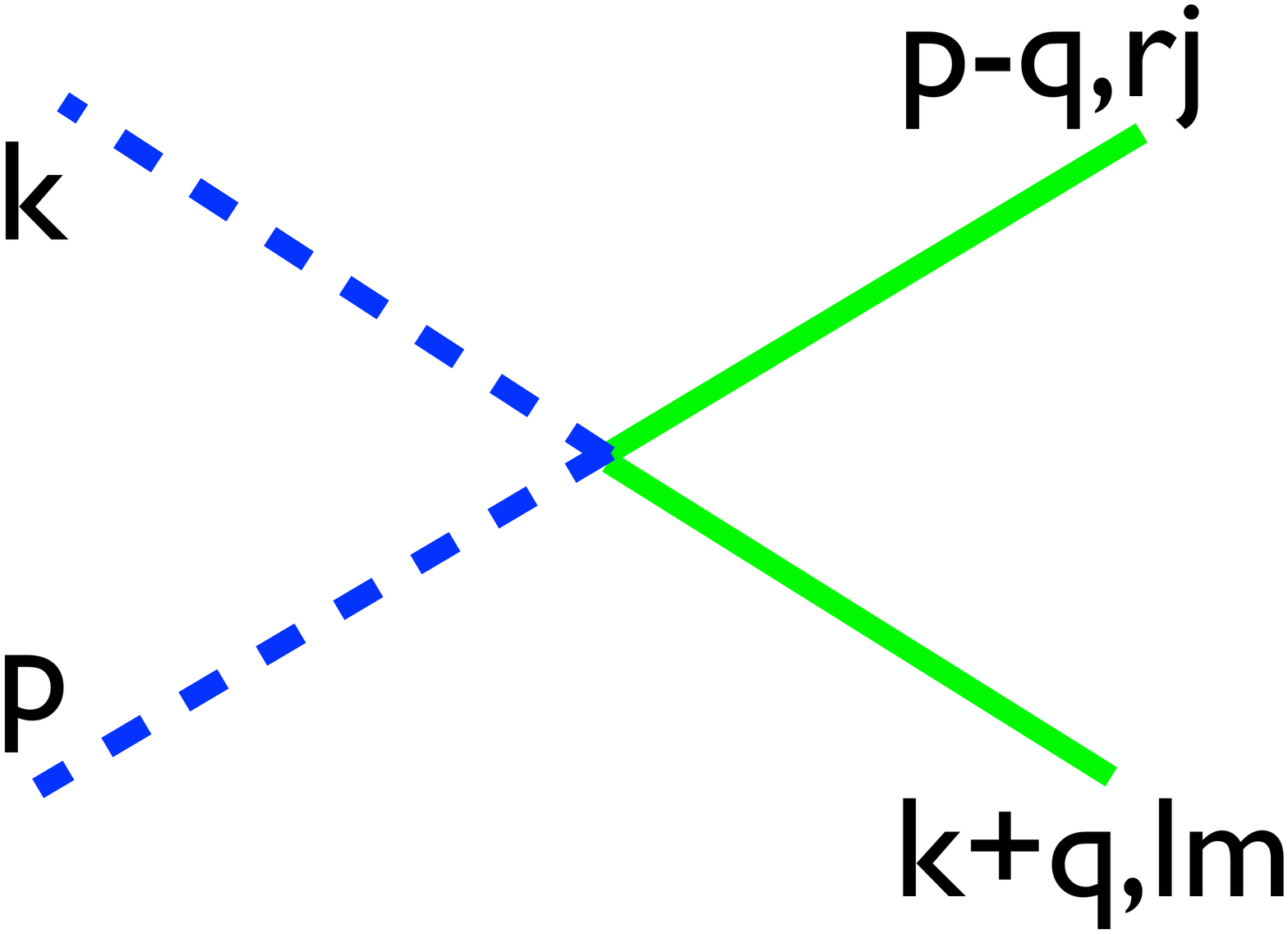}
\end{center}
\end{minipage}
&\rightarrow&\text{i}\frac{4 c_d}{\Lambda^2}\left[k^r p^l\delta^{jm}-\frac{1}{2}k^l p^m\delta^{rj}
-\frac{1}{8}p \cdot k{\cal Q}^{rjlm}+\Big(r\leftrightarrow j\,,l\leftrightarrow m\Big)\right],\nonumber\\
\begin{minipage}{2.0cm}
\begin{center}
\includegraphics[width=2.0cm]{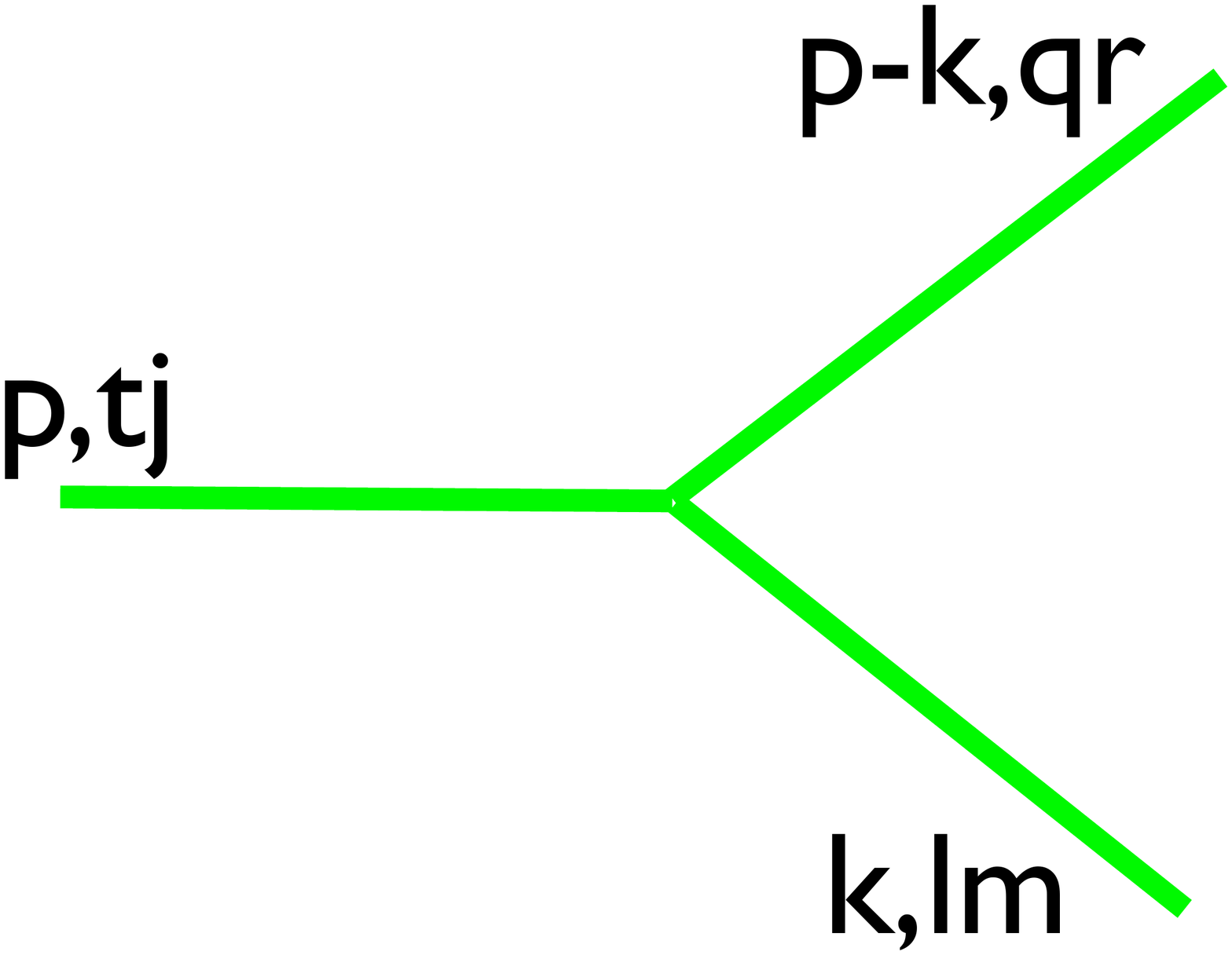}
\end{center}
\end{minipage}
&\rightarrow&\text{i} \frac{1}{8 \Lambda}
\Bigg\{(p-k) \cdot k\left(\frac{1}{2}\delta^{tr}{\cal I}^{lmjq}
-\frac{1}{4}\delta^{qr}{\cal I}^{tjlm}-\frac{1}{8}\delta^{tj}{\cal
  Q}^{qrlm}\right) + \\ 
&+& \frac{1}{4}(p-k)^t k^j{\cal Q}^{qrlm}
+\left[\left(\frac{1}{2}\delta^{tj}\delta^{mr}-\delta^{tr}\delta^{jm}\right)(p-k)^q
  k^l-\Big(l\leftrightarrow q\Big)\right]+\nonumber\\
&+&
\delta^{lm}\delta^{tr}(p-k)^q k^j-\delta^{tm}\delta^{qr}(p-k)^l k^j
+\Big(t\leftrightarrow j\,,l\leftrightarrow m\,,q\leftrightarrow r\Big)\Bigg\},
\nonumber
\\
\begin{minipage}{2.2cm}
\begin{center}
\includegraphics[width=1.8cm]{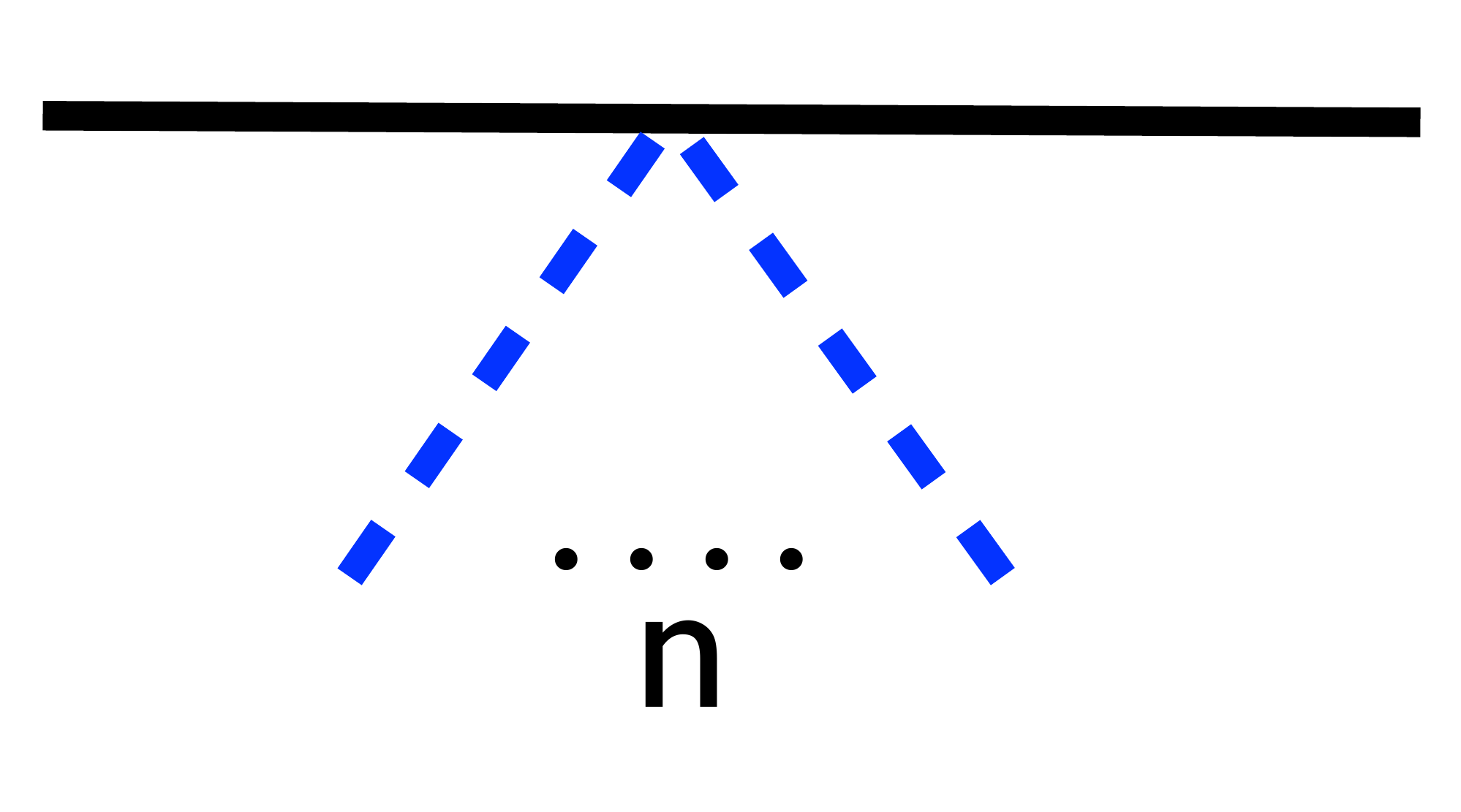}
\end{center}
\end{minipage}
&\rightarrow&-\frac{\text{i}}{n!  \Lambda^n}\nonumber
\end{eqnarray}
with ${\cal I}^{ijlm}\equiv\delta^{il}\delta^{jm}+\delta^{im}\delta^{jl}$ and ${\cal Q}^{ijlm}\equiv{\cal I}^{ijlm}-\delta^{ij}\delta^{lm}$.

Finally, the contribution of each amplitude to the two body Lagrangian ${\cal L}$ can be
derived from its Fourier transform, 
\begin{equation}
{\cal L}_a
\quad =  - \text{i}  \lim_{d \to 3} \ 
\int {{\rm d}^dp \over (2 \pi)^d} \ e^{\text{i} p \cdot r} 
\begin{minipage}{2.5cm}
\begin{center}
\includegraphics[width=2.0cm]{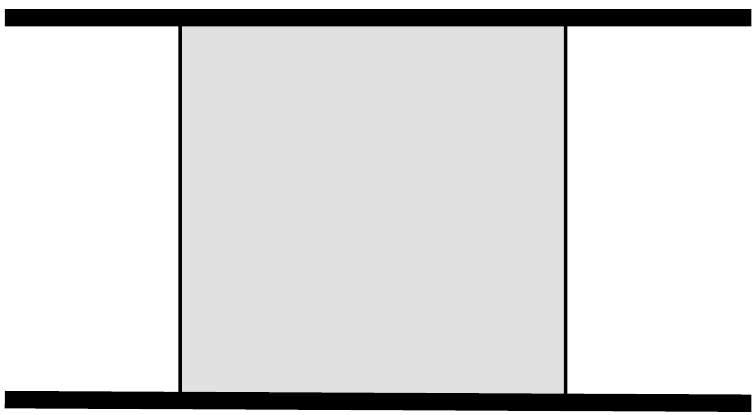}
\end{center}
\end{minipage}_a
\end{equation}
where the box diagram stands for the generic diagram $a=1,\ldots,50$ of
fig.~\ref{diaG5},
and $p$ is the momentum transfer of the source.

\section{Amplitudes and Feynman Integrals}
\label{sec:results}
In general, within the EFT approach, since the sources (black lines)
are static and do not propagate, any gravity-amplitude of order $G_N^\ell$ can be
mapped into an $(\ell-1)$-loop 2-point function with massless internal
lines and external momentum $p$, where $p^2 \equiv s \ne 0$,
\begin{equation}
\begin{minipage}{2.5cm}
\begin{center}
\includegraphics[width=2.0cm]{EFT_gravity}
\end{center}
\end{minipage}
\quad = \quad
\begin{minipage}{1.5cm}
\begin{center}
\includegraphics[width=1.0cm]{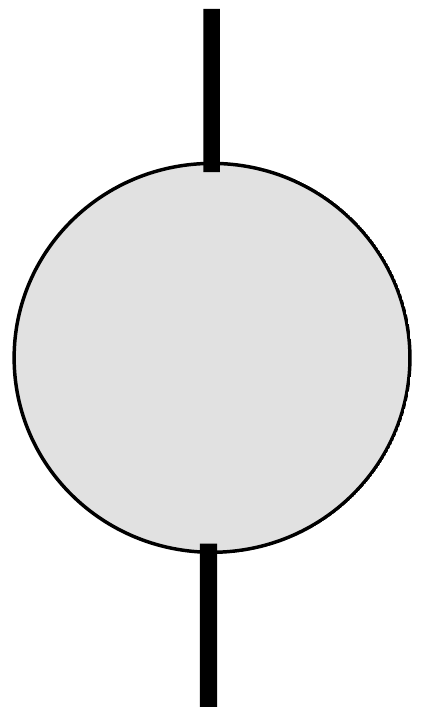} 
\end{center} 
\end{minipage} \ .
\end{equation}

\begin{figure}[t]
\begin{center}
\begin{minipage}{13cm}
\begin{minipage}{2cm}
\begin{center}
\includegraphics[width=1.5cm]{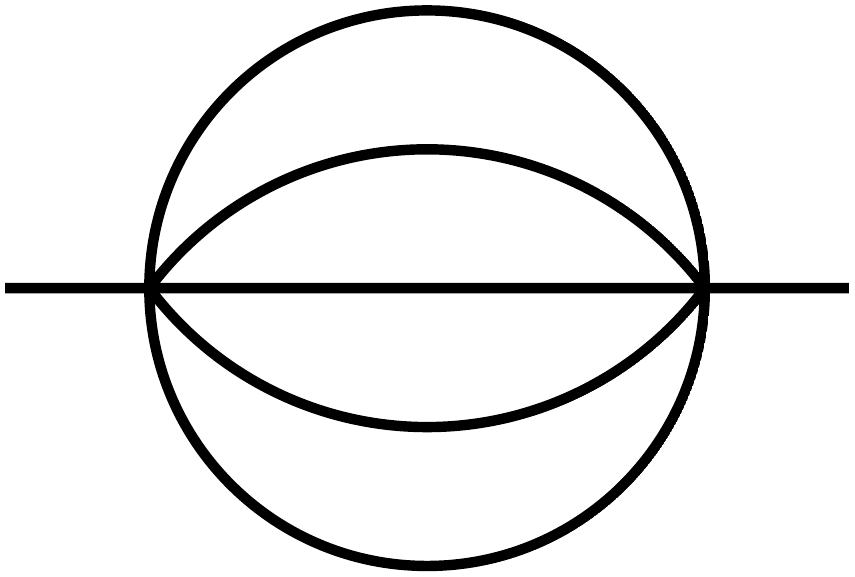}\\
{\tiny $T_1$}
\end{center}
\end{minipage}
\begin{minipage}{2cm}
\begin{center}
\includegraphics[width=1.5cm]{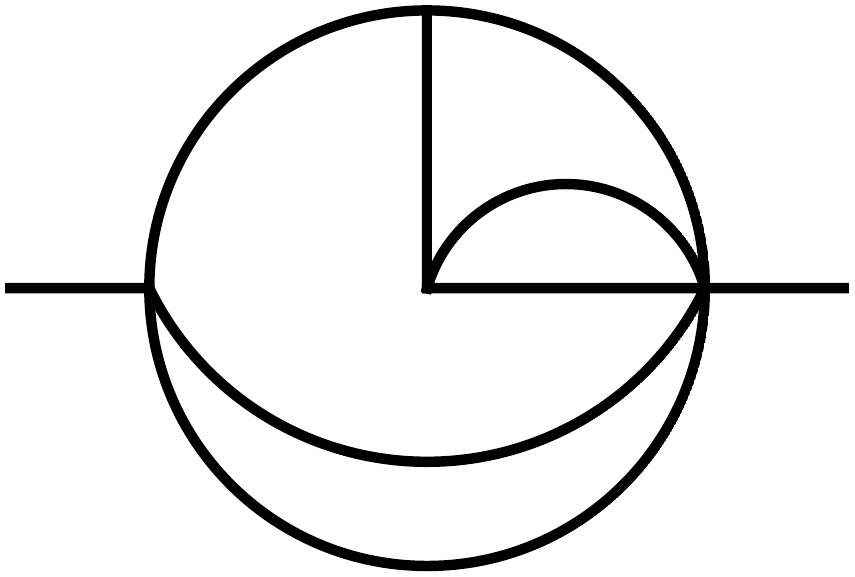}\\
{\tiny $T_2$}
\end{center}
\end{minipage}
\begin{minipage}{2cm}
\begin{center}
\includegraphics[width=1.5cm]{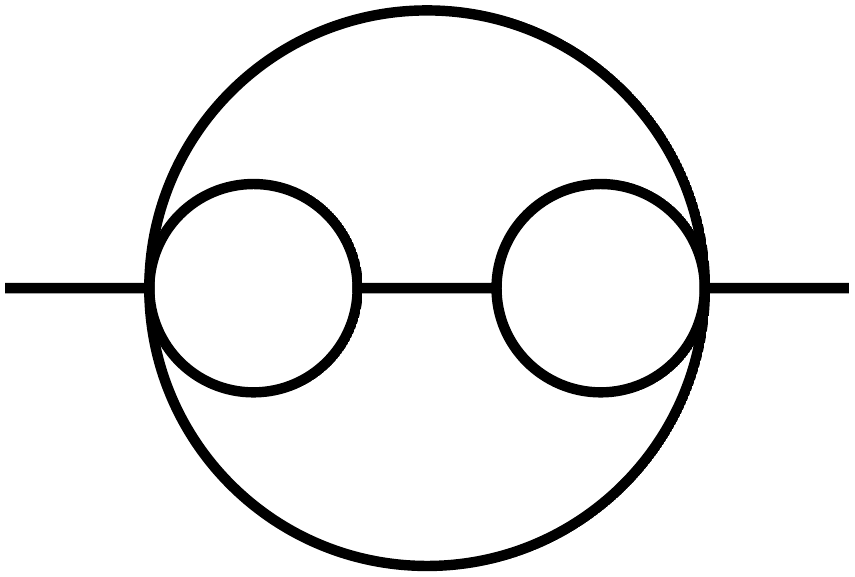}\\
{\tiny $T_3$}
\end{center}
\end{minipage}
\begin{minipage}{2cm}
\begin{center}
\includegraphics[width=1.5cm]{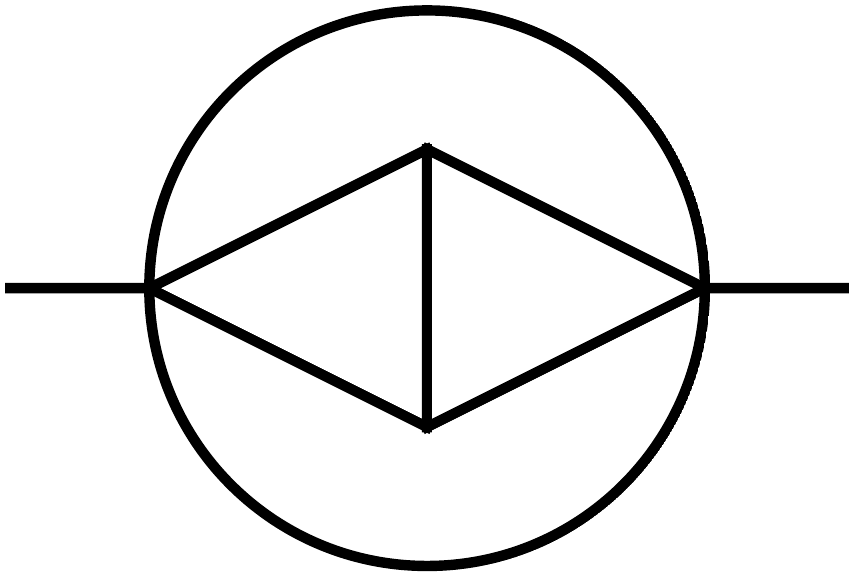}\\
{\tiny $T_4$}
\end{center}
\end{minipage}
\begin{minipage}{2cm}
\begin{center}
\includegraphics[width=1.5cm]{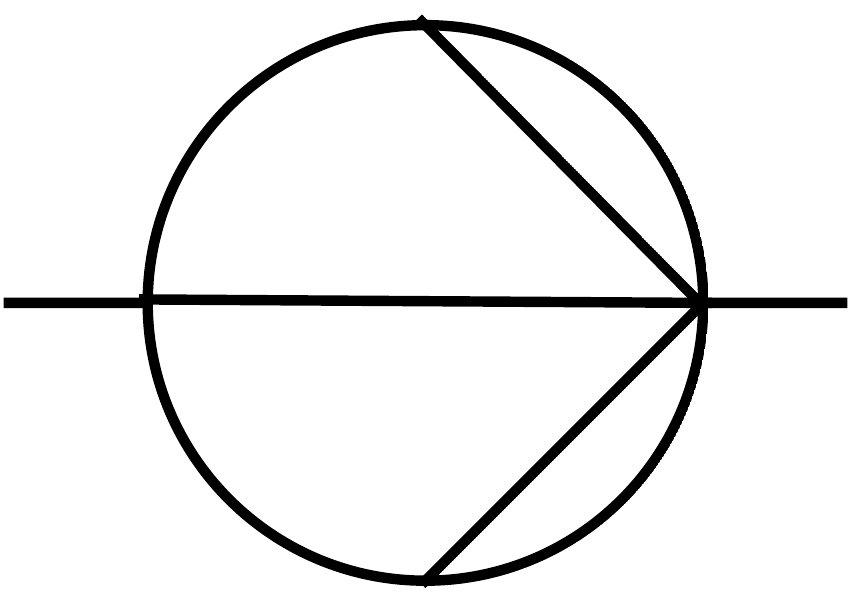}\\
{\tiny $26$}
\end{center}
\end{minipage}
\begin{minipage}{2cm}
\begin{center}
\includegraphics[width=1.5cm]{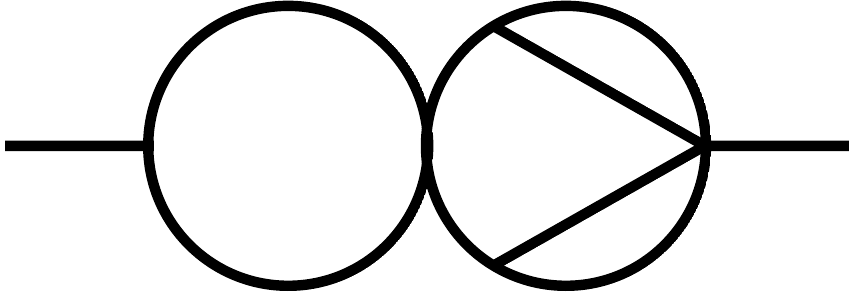}\\
{\tiny $27$}
\end{center}
\end{minipage}

\vspace*{0.5cm}

\begin{minipage}{2cm}
\begin{center}
\includegraphics[width=1.5cm]{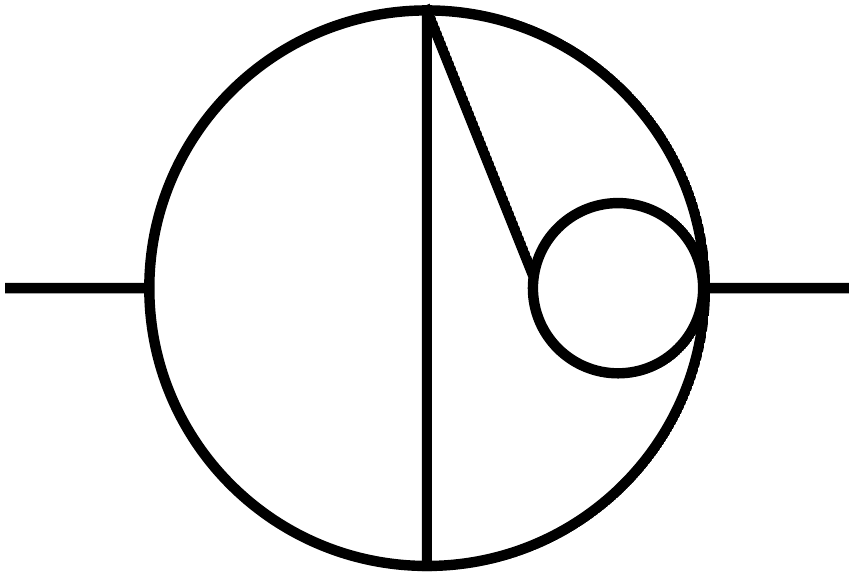}\\
{\tiny $28$}
\end{center}
\end{minipage}
\begin{minipage}{2cm}
\begin{center}
\includegraphics[width=1.5cm]{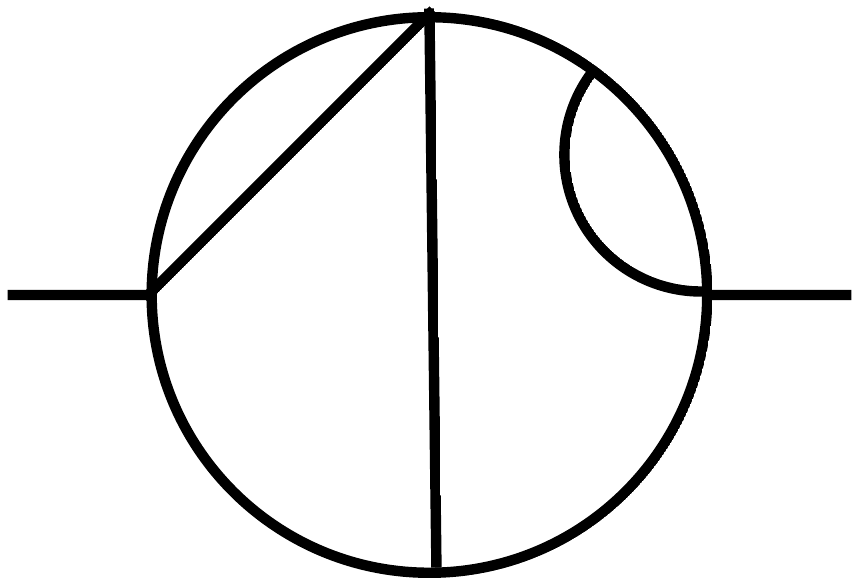}\\
{\tiny $29$}
\end{center}
\end{minipage}
\begin{minipage}{2cm}
\begin{center}
\includegraphics[width=1.5cm]{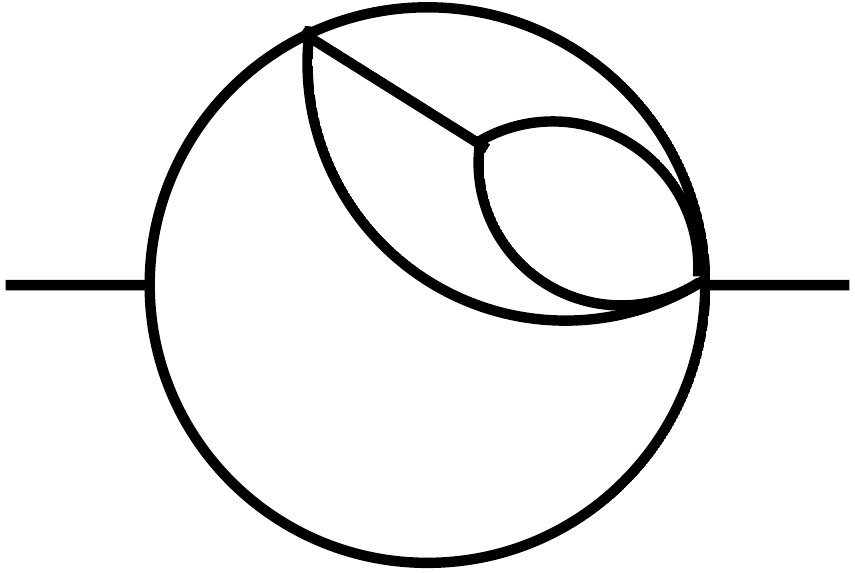}\\
{\tiny $30$}
\end{center}
\end{minipage}
\begin{minipage}{2cm}
\begin{center}
\includegraphics[width=1.5cm]{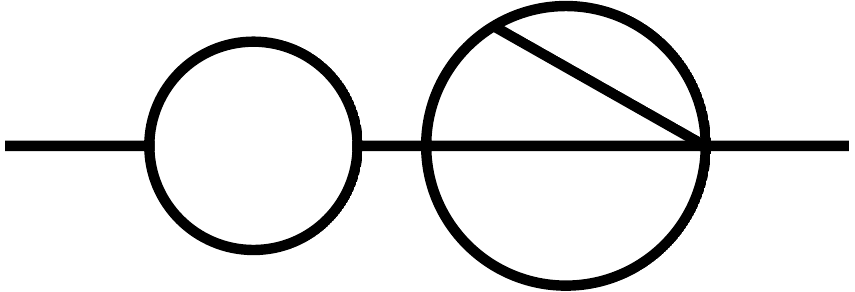}\\
{\tiny $31$}
\end{center}
\end{minipage}
\begin{minipage}{2cm}
\begin{center}
\includegraphics[width=1.5cm]{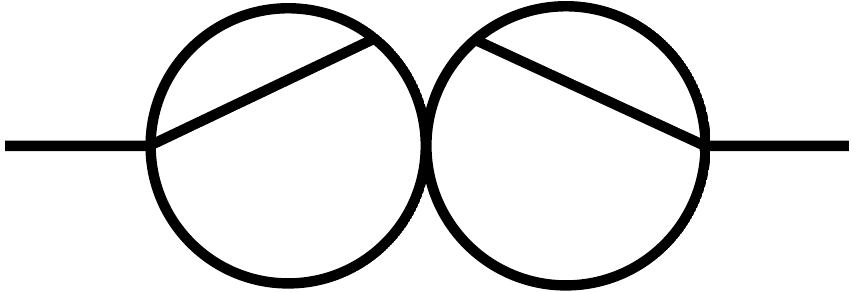}\\
{\tiny $32$}
\end{center}
\end{minipage}
\begin{minipage}{2cm}
\begin{center}
\includegraphics[width=1.5cm]{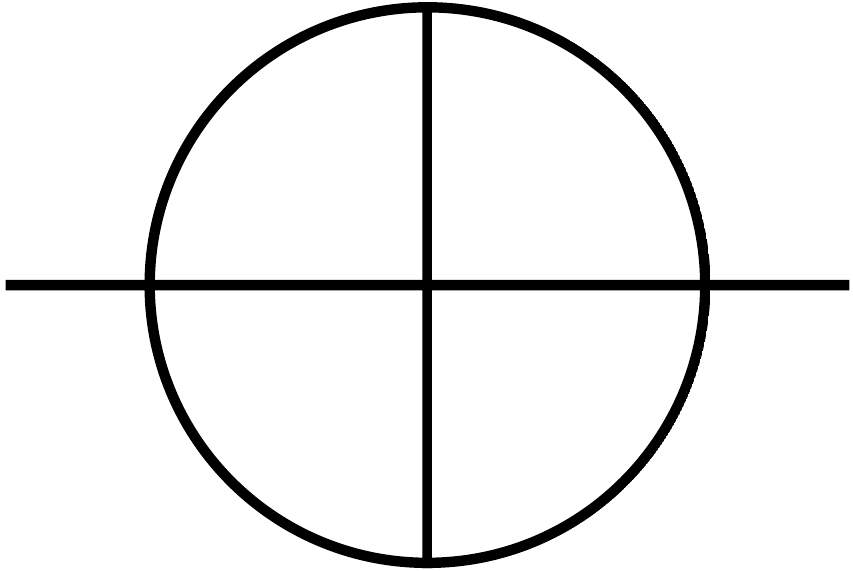}\\
{\tiny $33$}
\end{center}
\end{minipage}

\vspace*{0.5cm}

\begin{minipage}{2cm}
\begin{center}
\includegraphics[width=1.5cm]{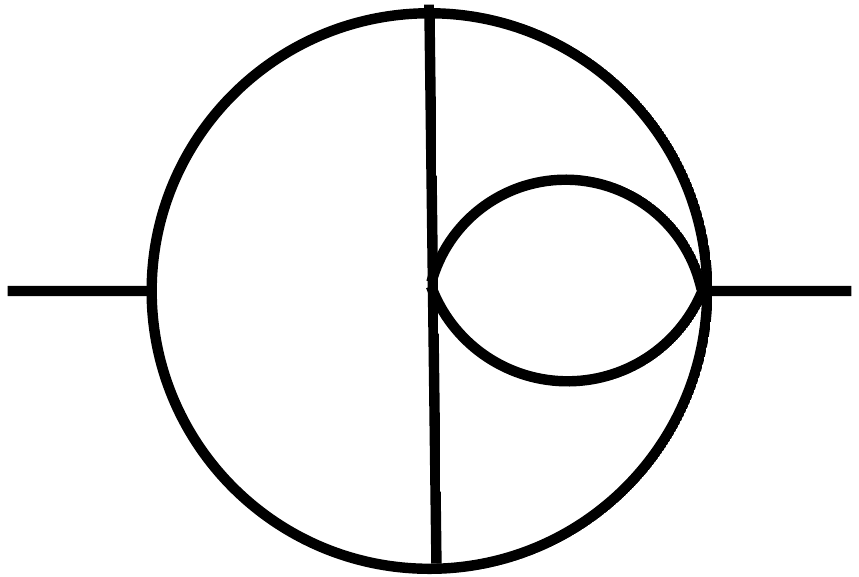}\\
{\tiny $34$}
\end{center}
\end{minipage}
\begin{minipage}{2cm}
\begin{center}
\includegraphics[width=1.5cm]{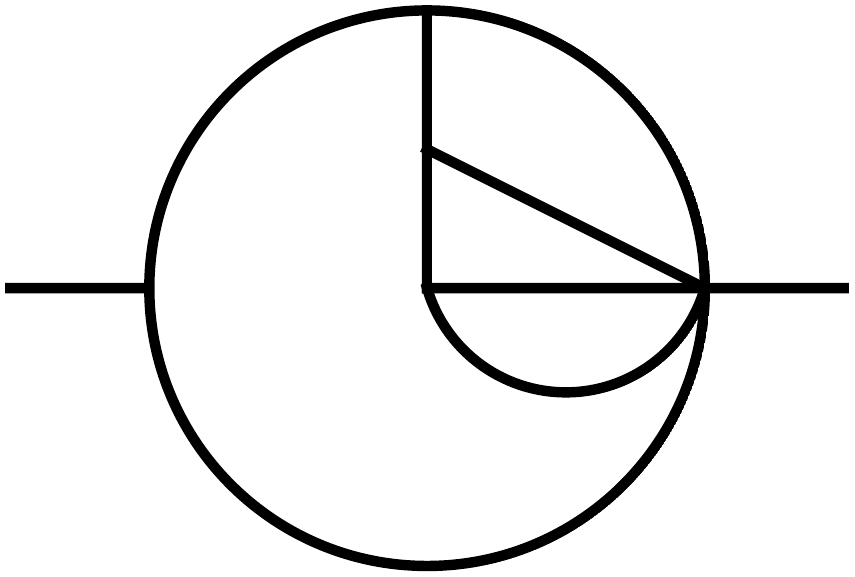}\\
{\tiny $35$}
\end{center}
\end{minipage}
\begin{minipage}{2cm}
\begin{center}
\includegraphics[width=1.5cm]{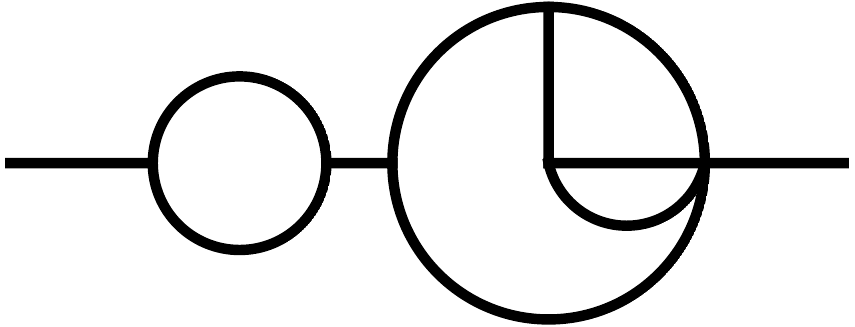}\\
{\tiny $36$}
\end{center}
\end{minipage}
\begin{minipage}{2cm}
\begin{center}
\includegraphics[width=1.5cm]{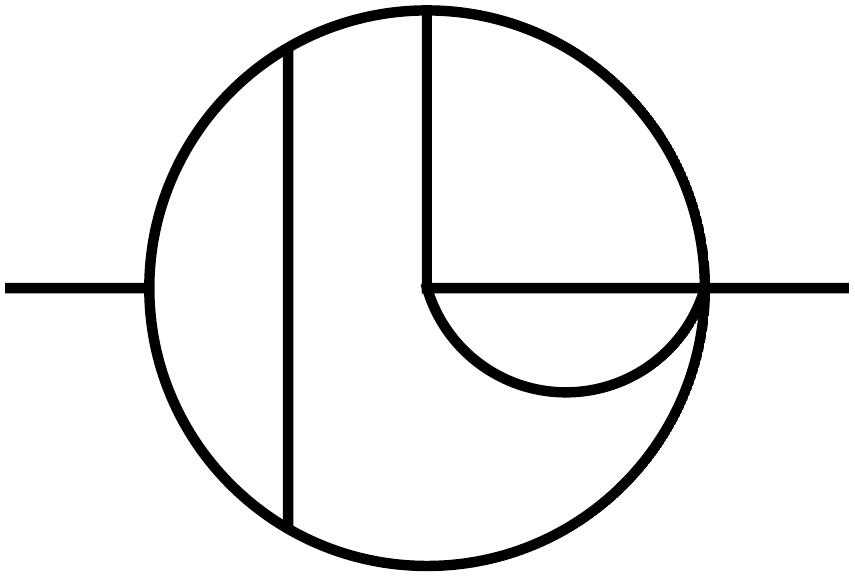}\\
{\tiny $37$}
\end{center}
\end{minipage}
\begin{minipage}{2cm}
\begin{center}
\includegraphics[width=1.5cm]{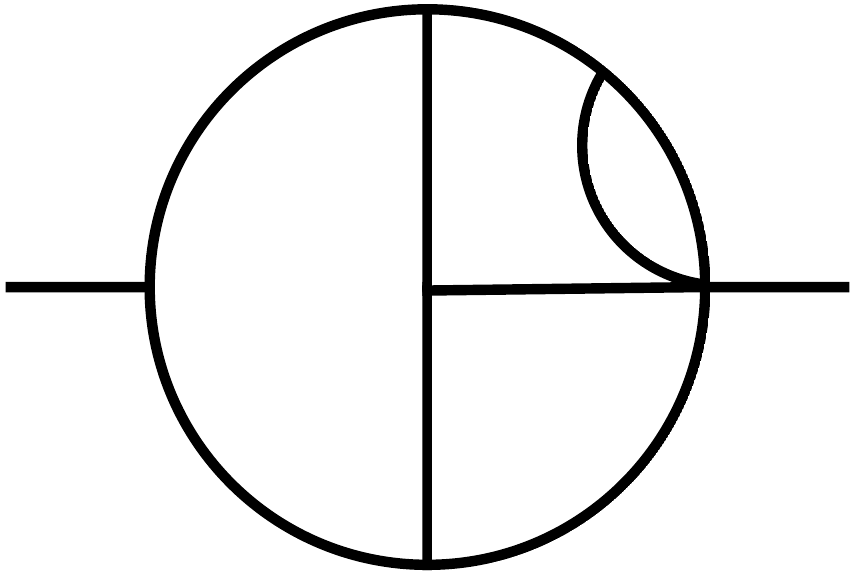}\\
{\tiny $38$}
\end{center}
\end{minipage}
\begin{minipage}{2cm}
\begin{center}
\includegraphics[width=1.5cm]{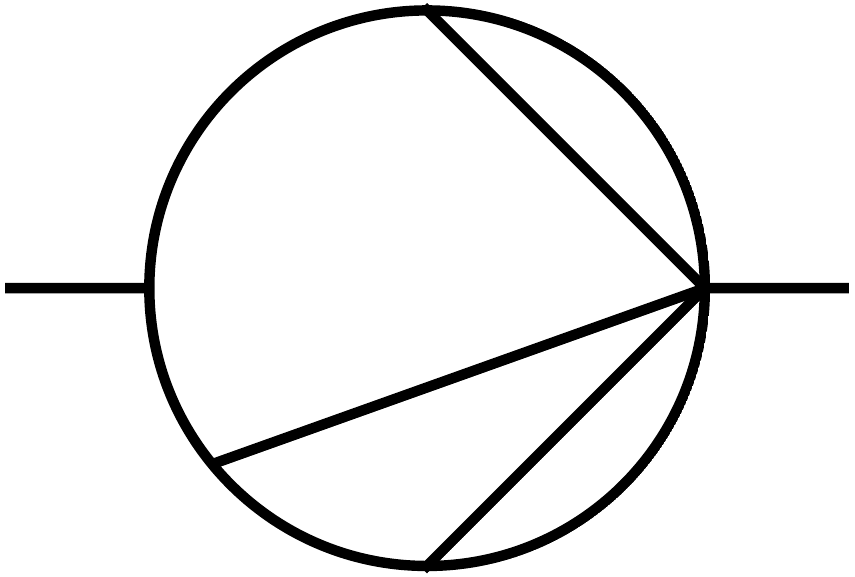}\\
{\tiny $39$}
\end{center}
\end{minipage}

\vspace*{0.5cm}

\begin{minipage}{2cm}
\begin{center}
\includegraphics[width=1.5cm]{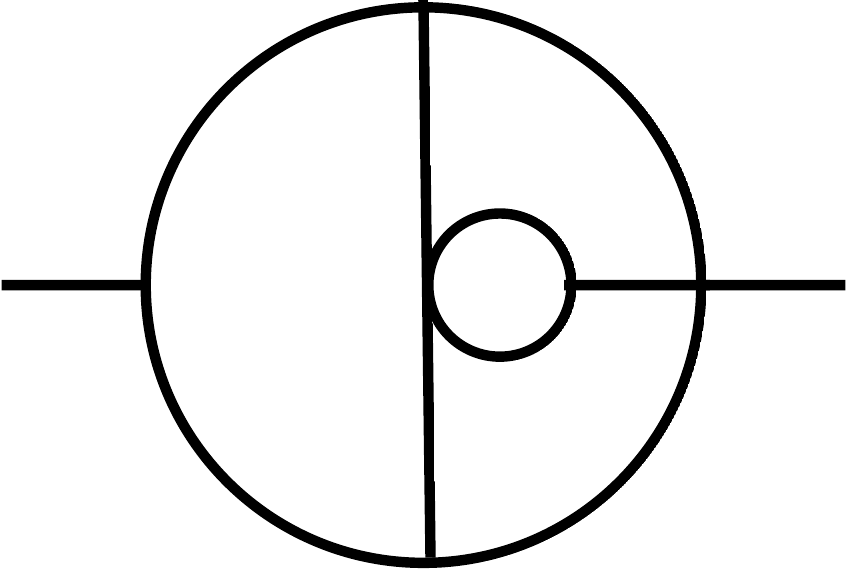}\\
{\tiny $40$}
\end{center}
\end{minipage}
\begin{minipage}{2cm}
\begin{center}
\includegraphics[width=1.5cm]{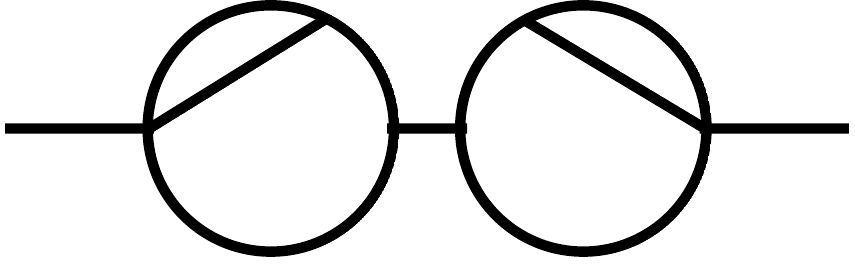}\\
{\tiny $41$}
\end{center}
\end{minipage}
\begin{minipage}{2cm}
\begin{center}
\includegraphics[width=1.5cm]{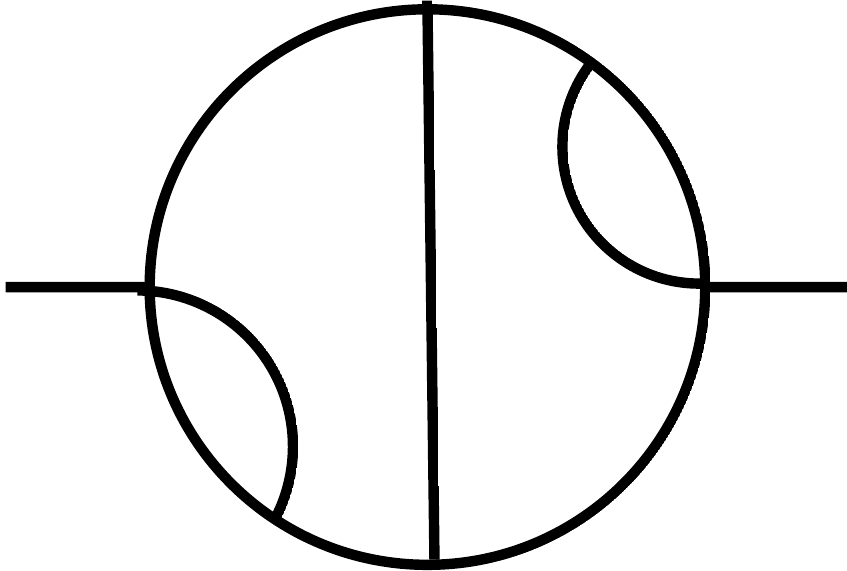}\\
{\tiny $42$}
\end{center}
\end{minipage}
\begin{minipage}{2cm}
\begin{center}
\includegraphics[width=1.5cm]{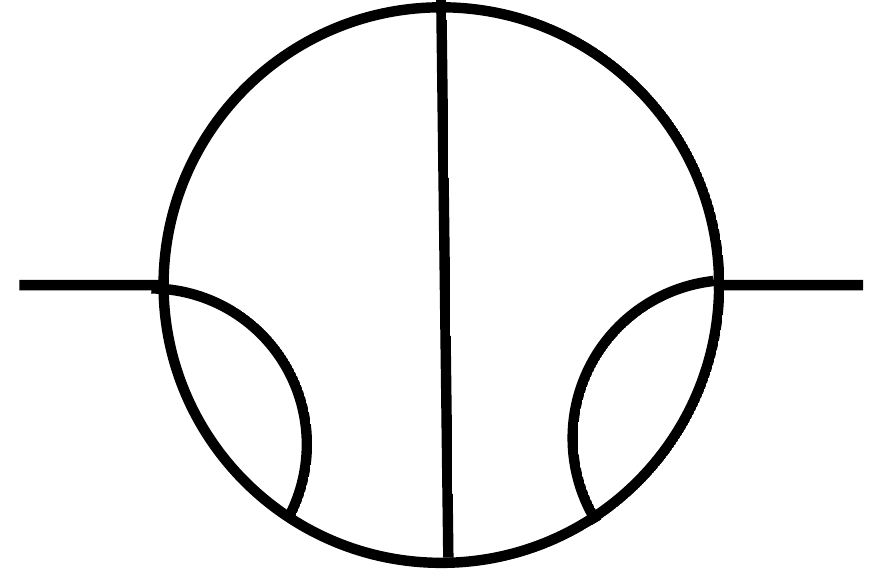}\\
{\tiny $43$}
\end{center}
\end{minipage}
\begin{minipage}{2cm}
\begin{center}
\includegraphics[width=1.5cm]{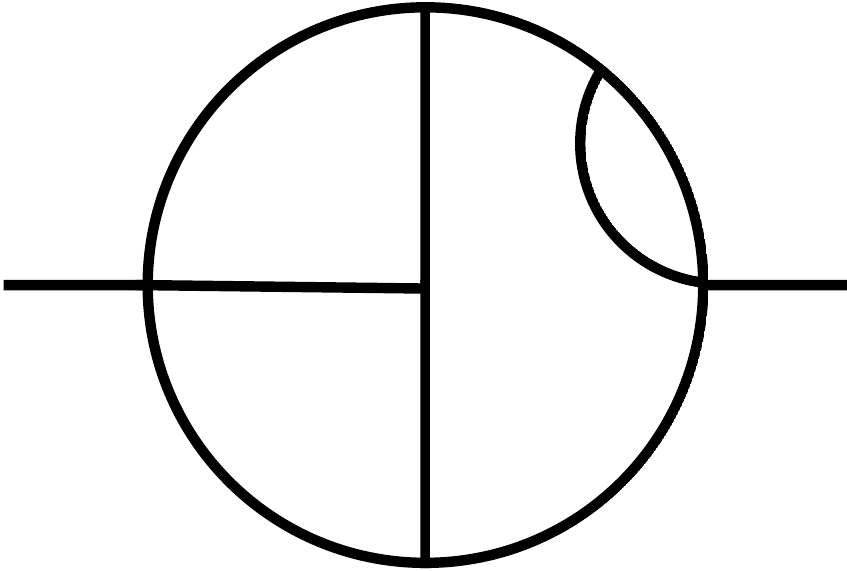}\\
{\tiny $44$}
\end{center}
\end{minipage}
\begin{minipage}{2cm}
\begin{center}
\includegraphics[width=1.5cm]{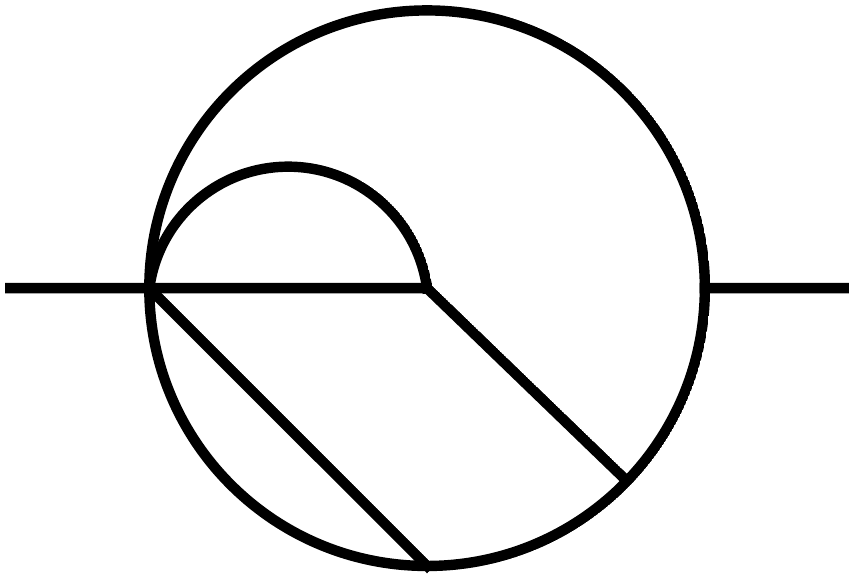}\\
{\tiny $45$}
\end{center}
\end{minipage}

\vspace*{0.5cm}

\begin{minipage}{2cm}
\begin{center}
\includegraphics[width=1.5cm]{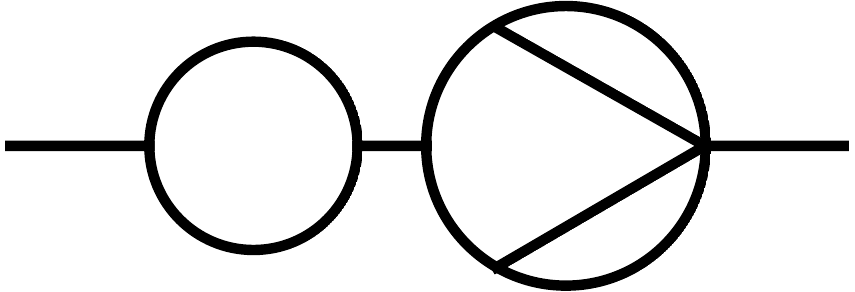}\\
{\tiny $46$}
\end{center}
\end{minipage}
\begin{minipage}{2cm}
\begin{center}
\includegraphics[width=1.5cm]{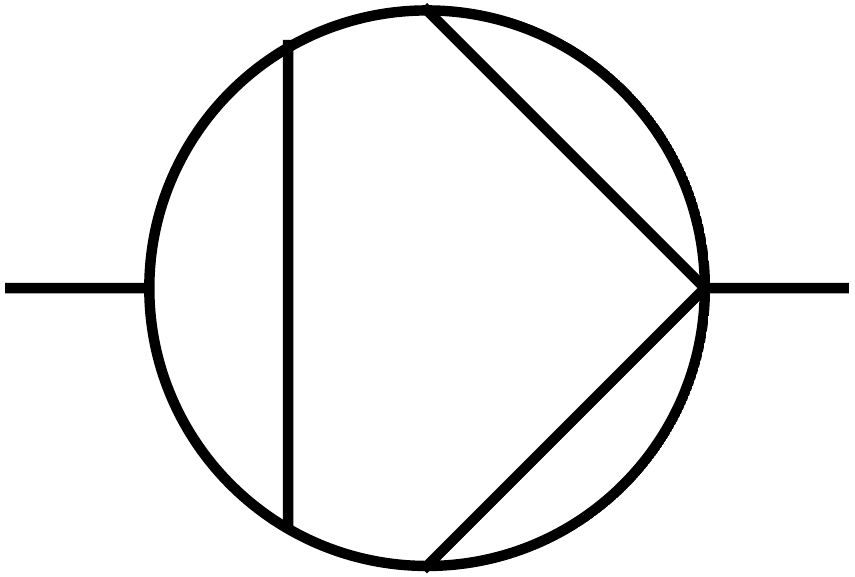}\\
{\tiny $47$}
\end{center}
\end{minipage}
\begin{minipage}{2cm}
\begin{center}
\includegraphics[width=1.5cm]{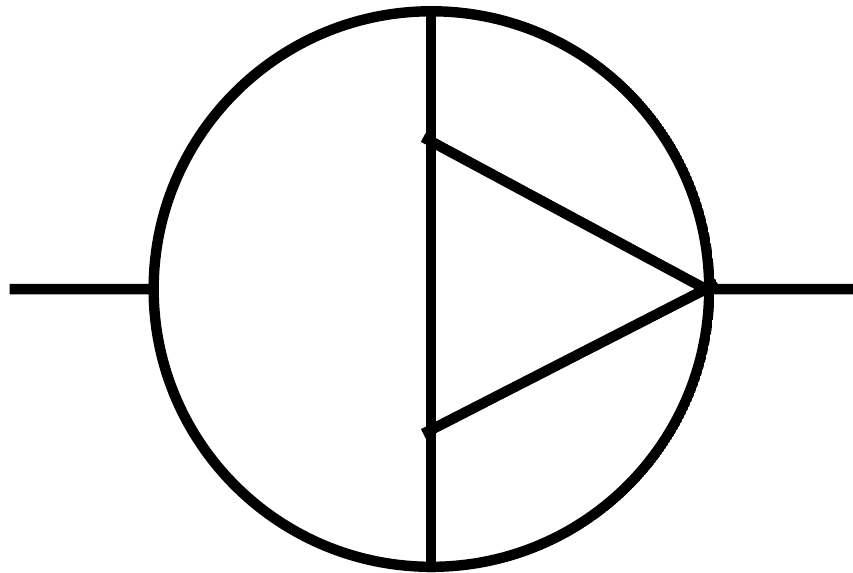}\\
{\tiny $48$}
\end{center}
\end{minipage}
\begin{minipage}{2cm}
\begin{center}
\includegraphics[width=1.5cm]{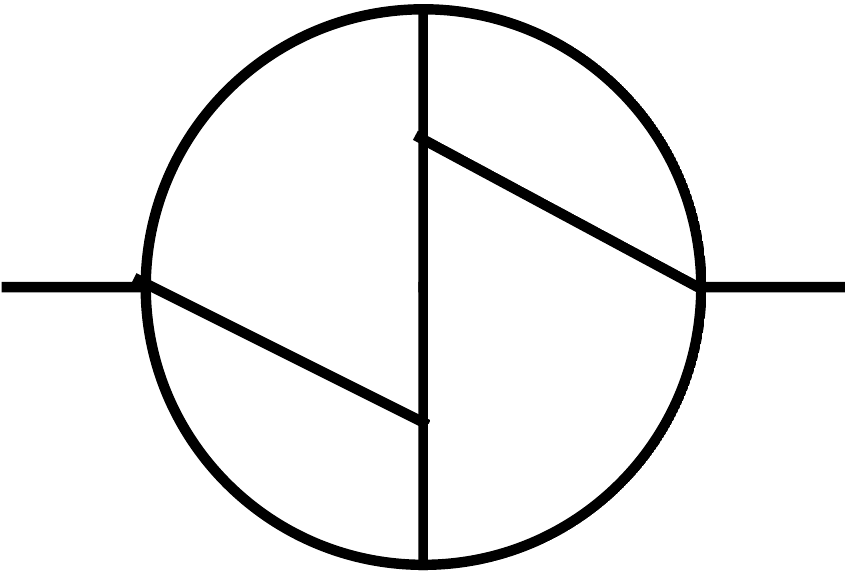}\\
{\tiny $49$}
\end{center}
\end{minipage}
\begin{minipage}{2cm}
\begin{center}
\includegraphics[width=1.5cm]{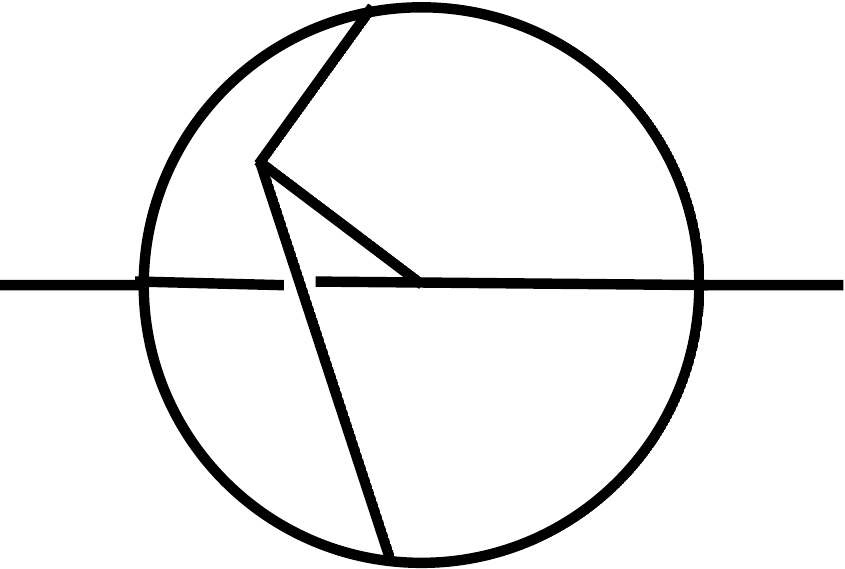}\\
{\tiny $50$}
\end{center}
\end{minipage}
\caption{\label{fig:topologies} Four-loop 2-point  topologies corresponding to
  the diagrams in fig.\ref{diaG5}.}
\end{minipage}
\end{center}
\end{figure}

\noindent
Accordingly, the 50 diagrams in fig.\ref{diaG5} can be mapped onto the
29 topologies of fig.\ref{fig:topologies},

where the sets 
$T_1 = \{ 1,2,3,4,5,6\}$, 
$T_2 = \{7,8,10,11,14,16,17,20,21,25\}$,
$T_3 = \{ 9,12,13,22\}$, 
$T_4 = \{15, 18,19,23,24\}$, 
collect the diagrams that share the same topology.
For instance, the diagrams 1 to 6 of fig.\ref{diaG5} correspond to integrals which have
the same five denominators of the graph indicated by $T_1$ in 
fig.\ref{fig:topologies},
but different numerators, due to the different terms 
associated to 1,2,3 or 4 $\phi$ emission or absorption from the massive
particle.

The representation of the gravity-amplitudes as 4-loop 2-point integrals yields the
possibility of evaluating the latter by means of by-now standard multi-loop
techniques based on integration-by-parts 
identities (IBPs)~\cite{Tkachov:1981wb,Chetyrkin:1981qh}. 

Accordingly, we collect the 50 amplitudes of fig.\ref{diaG5} in two
sets, $\mathcal{A}_I=\{1:28,31,32,35:37,39,41,45:47\}$ and 
$\mathcal{A}_{II}=\{29,30,33,34,38,40,42,43,44,48,49,50\}$, and
address their computation separately.

\begin{figure}[t]
\begin{center}
\begin{minipage}{3cm}
\begin{center}
\includegraphics[width=2.5cm]{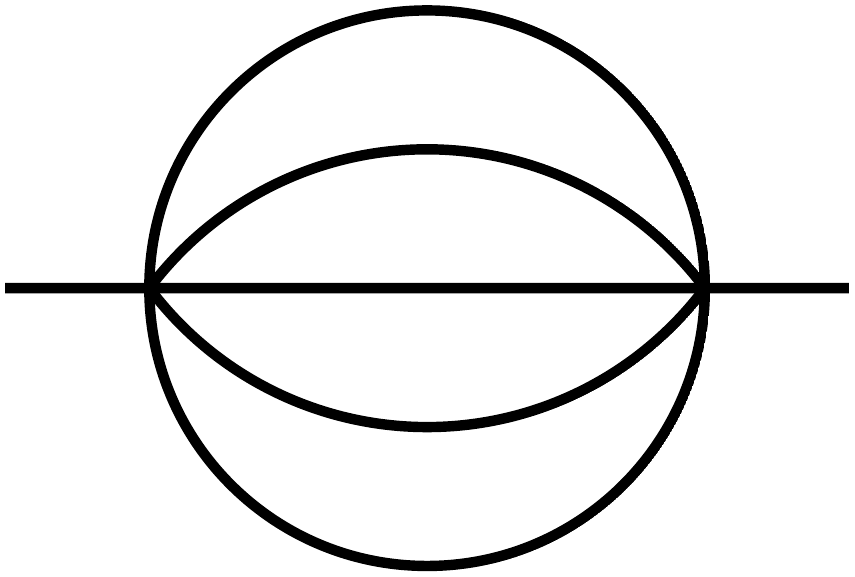}\\
$\mathcal{M}_{0,1}$
\end{center}
\end{minipage}
\begin{minipage}{3cm}
\begin{center}
\includegraphics[width=2.5cm]{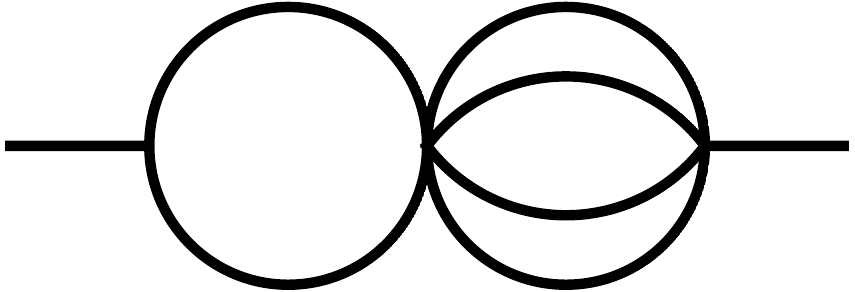}\\
$\mathcal{M}_{1,1}$
\end{center}
\end{minipage}
\begin{minipage}{3cm}
\begin{center}
\includegraphics[width=2.5cm]{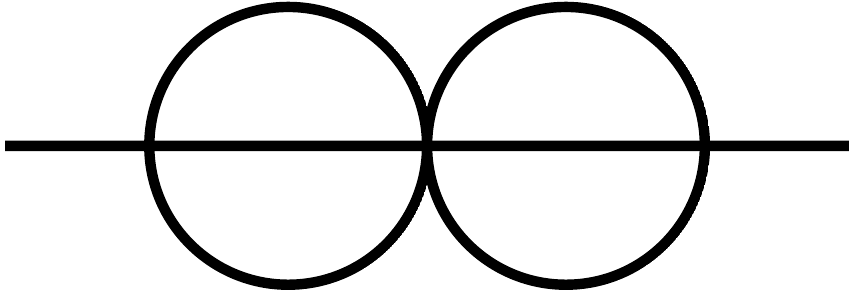}\\
$\mathcal{M}_{1,2}$
\end{center}
\end{minipage}\\[0.3cm]
\begin{minipage}{3cm}
\begin{center}
\includegraphics[width=2.5cm]{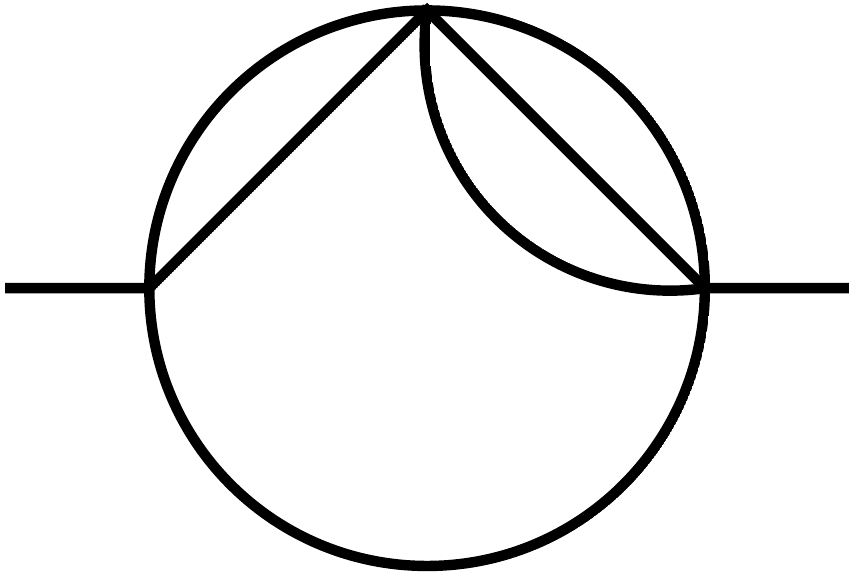}\\
$\mathcal{M}_{1,3}$
\end{center}
\end{minipage}
\begin{minipage}{3cm}
\begin{center}
\includegraphics[width=2.5cm]{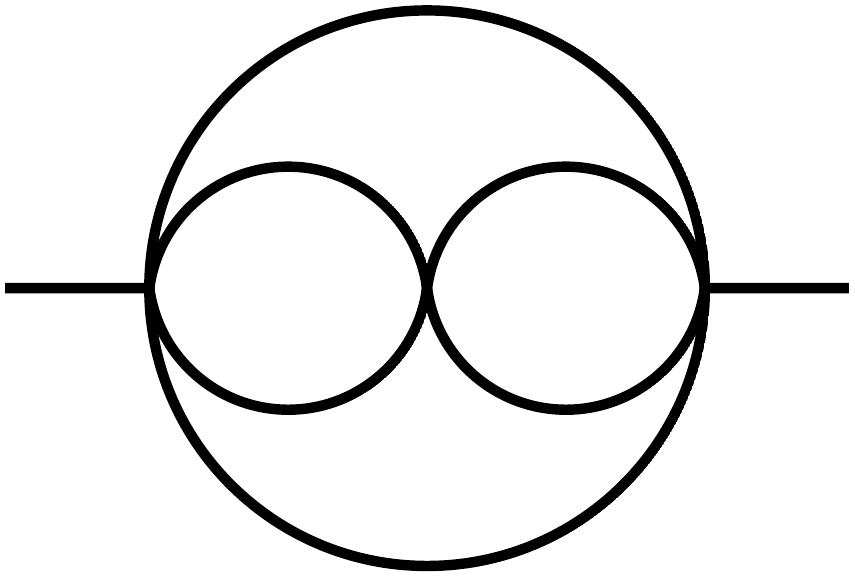}\\
$\mathcal{M}_{1,4}$
\end{center}
\end{minipage}
\begin{minipage}{3cm}
\begin{center}
\includegraphics[width=2.5cm]{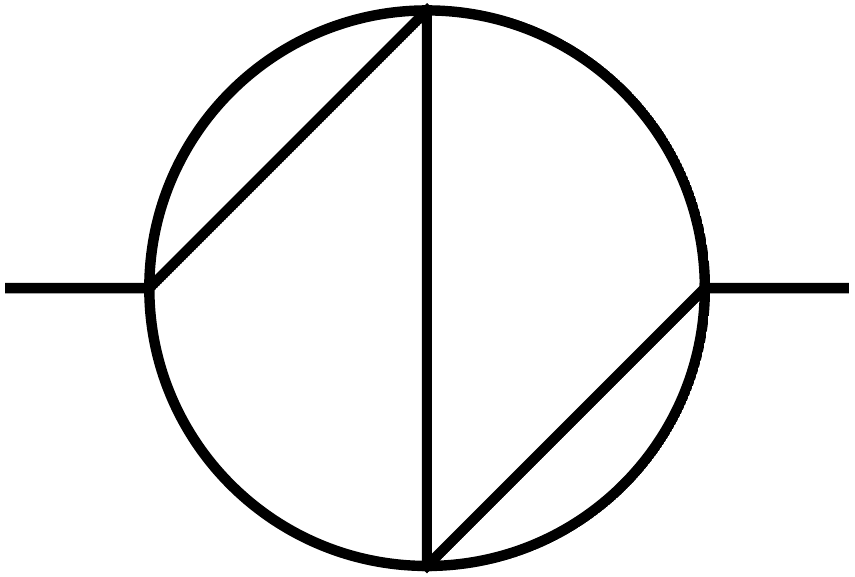}\\
$\mathcal{M}_{2,2}$
\end{center}
\end{minipage}
\begin{minipage}{3cm}
\begin{center}
\includegraphics[width=2.5cm]{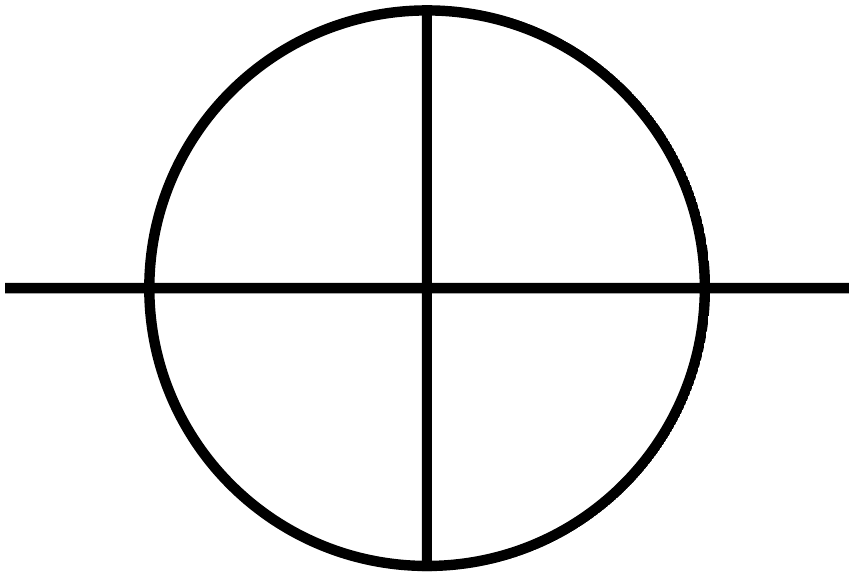}\\
$\mathcal{M}_{3,6}$
\end{center}
\end{minipage}
\end{center}
\caption{\label{fig:masters}The
  master integrals which appear in the calculation of the amplitudes in
  the set $\mathcal{A}_{II}$. The names of the diagrams follow
  refs.~\cite{Baikov:2010hf,Lee:2011jt,Lee:2015eva}.}
\end{figure}

The set  $\mathcal{A}_I$ contains diagrams with a simpler internal
structure, and they have been computed
by using the kite rule~\cite{Tkachov:1981wb,Chetyrkin:1981qh}
\begin{equation}
{(4-d) \over 2} \ 
\begin{minipage}{2.5cm}
\includegraphics[width=2.3cm]{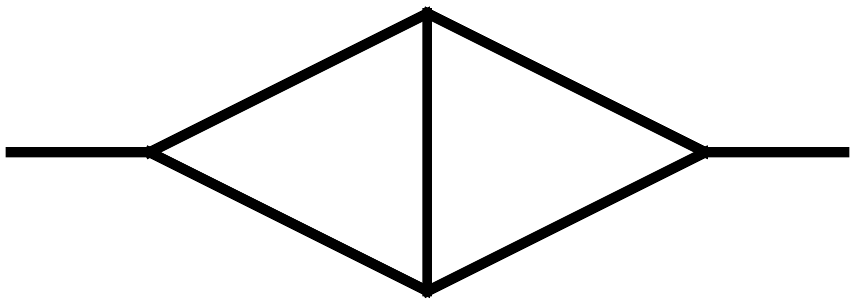}
\end{minipage}
= \ 
\begin{minipage}{1.9cm}
\includegraphics[width=1.8cm]{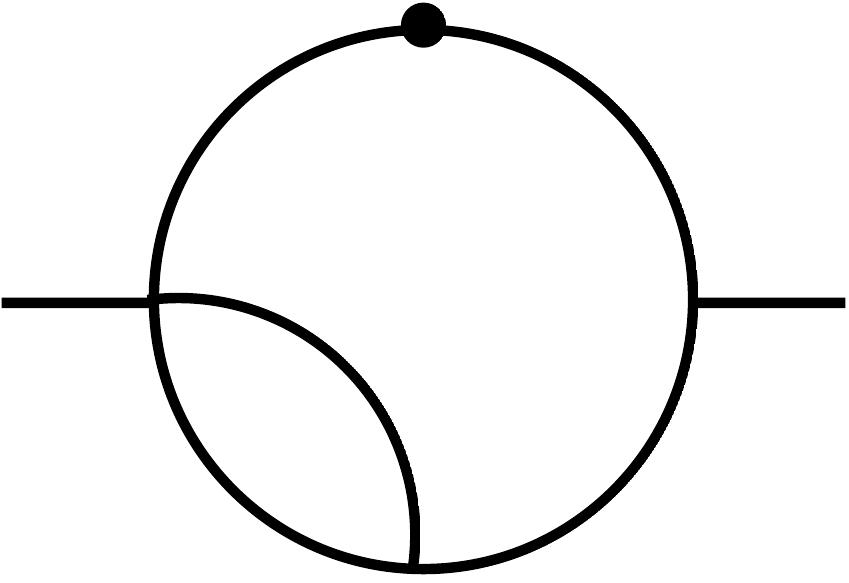}
\end{minipage}
- \ 
\begin{minipage}{2cm}
\includegraphics[width=2.0cm]{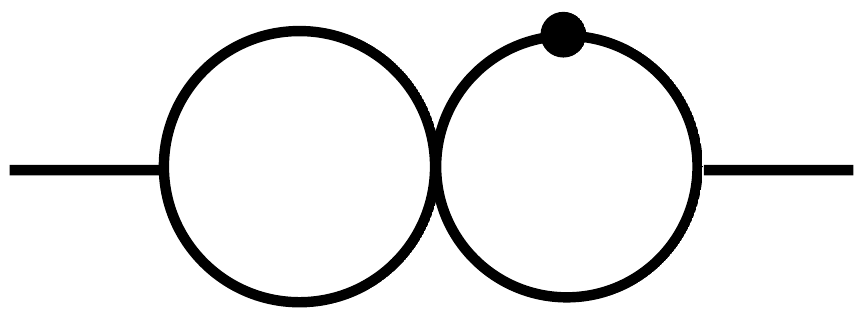}
\end{minipage}
\ , 
\end{equation}

where the dots stand for squared denominators,
and by using the standard identity holding for 2-point 1-loop graphs, 
\begin{eqnarray}
\label{1loop}
\int\!\!\!\!\frac{{\rm d}^dk}{(2\pi)^d}\frac 1{k^{2a}(p-k)^{2b}}&=&
\begin{minipage}{1.6cm}
\begin{center}
\includegraphics[width=1.5cm]{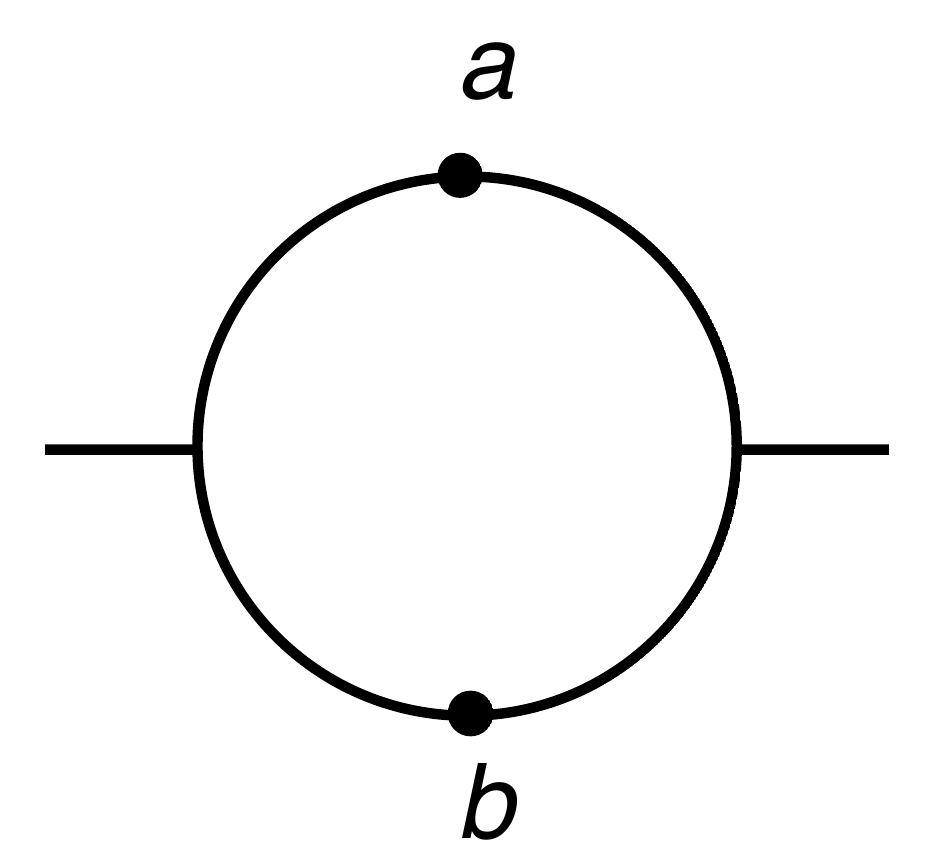}
\end{center}
\end{minipage}=
\frac{\left(p^2\right)^{d/2-a-b}}{(4\pi)^{d/2}}
\frac{\Gamma(d/2-a)\Gamma(d/2-b)\Gamma(a+b-d/2)}
{\Gamma(a)\Gamma(b)\Gamma(d-a-b)}\,,
\end{eqnarray}
where $a$ and $b$ are generic denominators' powers.
Alternatively we also performed an IBP-reduction using  the program
{\tt{Reduze}}~\cite{Studerus:2009ye,vonManteuffel:2012np}, 
identifying 5 master integrals (MIs), namely $\mathcal{M}_{0,1}$, $\mathcal{M}_{1,1}$,
$\mathcal{M}_{1,2}$, $\mathcal{M}_{1,3}$, $\mathcal{M}_{1,4}$ of fig.~\ref{fig:masters}.
Both strategies gave the same results.

The amplitudes $\mathcal{A}_{II} $, instead, have a less trivial
internal structure. By means of IBPs, they have been systematically reduced
to linear combinations of 7 MIs, all shown in fig.~\ref{fig:masters}.
In this case, the reduction to MIs has been performed in two ways, 
by an in-house implementation of Laporta's algorithm which is based on
{\tt{Form}}~\cite{Vermaseren:2000nd,Vermaseren:2002rp,Tentyukov:2006ys},
as well as by means of 
{\tt{Reduze}}.

The 4-loop MIs in fig.~\ref{fig:masters} can be considered as a complete set of independent integrals,
such that any amplitude of the sets $\mathcal{A}_{I} $ and
$\mathcal{A}_{II}$ can be written as a linear combination of them.
The results of the 4-loop MIs are well-known in $d=4+\varepsilon$ euclidean space-time dimensions since
long~\cite{Baikov:2010hf,Lee:2011jt}, while the values around $d=3+\varepsilon$ 
of $\mathcal{M}_{2,2}$, $\mathcal{M}_{3,6}$ became
available more recently \cite{Lee:2015eva}.
In particular, $\mathcal{M}_{0,1}$, $\mathcal{M}_{1,1}$,
$\mathcal{M}_{1,2}$, $\mathcal{M}_{1,3}$, $\mathcal{M}_{1,4}$ can be
computed in a straightforward way by means of eq.~(\ref{1loop}), and
admit closed analytic expressions, exact in $d$, which can be expanded
in Laurent series in $\varepsilon$ around $d=3$.  
The series expansions of $\mathcal{M}_{2,2}$ and $\mathcal{M}_{3,6}$ 
were first obtained numerically in ref.~\cite{Lee:2015eva} by using the {\it difference equations method}, 
exploiting the fact that dimensionally regulated Feynman integrals obey dimensional recurrence
relations~\cite{Milne-Thomson:1933:CFD,Derkachov:1990osa,Tarasov:1996br,Laporta:2001dd,Lee:2009dh}.
For instance, owing to IBPs, $\mathcal{M}_{3,6}$ is solution of
the following recursive formula,
\begin{eqnarray}
{1 \over (4 \pi)^4} \cdot 
\begin{minipage}{2.0cm}
\begin{center}
\includegraphics[width=2.0cm]{1040802}
\end{center}
\end{minipage}\Bigg|_{d-2} 
&=&
a_1
\begin{minipage}{2.0cm}
\begin{center}
\includegraphics[width=2.0cm]{1040802}
\end{center}
\end{minipage}
+
a_2
\begin{minipage}{2.0cm}
\begin{center}
\includegraphics[width=1.9cm]{1040603}
\end{center}
\end{minipage}
+
a_3
\begin{minipage}{2.0cm}
\begin{center}
\includegraphics[width=1.9cm]{1040602}
\end{center}
\end{minipage}
+
\nonumber \\
& & 
+
a_4
\begin{minipage}{2.0cm}
\begin{center}
\includegraphics[width=1.9cm]{1040607}
\end{center}
\end{minipage}
+
a_5
\begin{minipage}{2.0cm}
\begin{center}
\includegraphics[width=1.9cm]{1040501}
\end{center}
\end{minipage}
\ .
\label{eq:recrel:M36}
\end{eqnarray}
with 
\begin{eqnarray}
a_1 &=&{5\*(d-3)\*(d-4)^2\*(5-d)\*(5\*d-26)\*(5\*d-24)\*(5\*d-22)\*(5\*d-18) 
      \over 3\*(d-6)^2\*(3\*d-16)\*(3\*d-14)\*s^4},\\
a_2 &=&80\*(d-3)^3\*(2\*d-7)\*(5\*d-26)\*(5\*d-24)\*(5\*d-22)\*(5\*d-18)\*(5\*d-16)\* 
\times\nonumber\\&&\qquad
 {(14-5\*d)\*(63872 - 40162\*d + 8403\*d^2 - 585\*d^3) \over
 9\*(d-6)^2\*(d-4)^2\*(3\*d-16)^2\*(3\*d-14)^2\*(3\*d-10)\*s^6},\\
a_3 &=&
40\*(d-3)^2\*(8-3\*d)\*(5\*d-26)\*(5\*d-24)\*(5\*d-22)\*(5\*d-18)\*
\times\nonumber\\&&\qquad
{(5\*d-16)\*(5\*d-14)\*(7\*d-32)
\over 3\*(d-6)^2\*(d-4)^2\*(3\*d-16)\*(3\*d-14)\*s^6},\\
a_4 &=&(d-3)^2\*(3\*d-10)^2\*(3\*d-8)^2\*
\times\nonumber\\&&\qquad
{2897664 - 2445164\*d + 772948\*d^2 -  108475\*d^3 + 5702\*d^4 
\over 3\*(d-6)^2\*(d-4)^2\*(3\*d-16)\*(3\*d-14)\*s^6},\\
a_5 &=&
20\*(d-3)\*(2\*d-7)\*(2\*d-5)\*(5\*d-26)\*(5\*d-24)\*
\times\nonumber\\&&\qquad
(5\*d-22)\*(5\*d-18)\*(5\*d-16)\*(5\*d-14)\*(5\*d-12)\*
\times\nonumber\\&&\qquad
{(1972736-1666418\*d+527297\*d^2-74070\*d^3+3897\*d^4)
\over 9\*(d-6)^2\*(d-5)\*(d-4)^3\*(3\*d-16)^2\*(3\*d-14)^2\*s^7} \ , 
\end{eqnarray}
which links $M_{3,6}$ in $d-2$ dimensions (on the l.h.s.) to 
$M_{3,6}$ in $d$ dimension, and to other MIs belonging to
subtopologies, also defined in $d$ dimensions (on the r.h.s). 
The MIs belonging to subtopologies have to be considered as the non-homogeneous term of the
dimensional recurrence relation: they are known terms in a
bottom-up approach (where simpler integrals, with less denominators,
are computed first)
\footnote{ The dimensional recurrence (\ref{eq:recrel:M36}) implies that
  ${\cal M}_{3,6}(d=3+\varepsilon) \equiv \sum_{k=-2}^\infty {\cal
    M}_{3,6}(3,k) \varepsilon^k$ can be obtained from the knowledge of
  the MIs on the r.h.s., ${\cal M}_{i,j}(d=5+\varepsilon) \equiv
  \sum_{k=-2}^\infty {\cal M}_{i,j}(5,k) \varepsilon^k $.  It is
  interesting to notice that in eq.~(\ref{eq:recrel:M36}) the
  coefficient $a_1$ is proportional to $(d-5)$.  Therefore, by expanding
  both sides of the equation in a Laurent series, the Laurent
  coefficient ${\cal M}_{3,6}(3,k)$ gets a contribution from ${\cal
    M}_{3,6}(5,k-1)$ and from the Laurent coefficients of the other MIs
  at $d=5$.  In particular, the coefficient of the double pole ${\cal
    M}_{3,6}(3,-2)$ is completely determined by the series expansions of
  the MIs of the subtopologies only, because when $k=-2$, ${\cal
    M}_{3,6}(d=5+\varepsilon)$ does not give any contribution.}.

The solving strategy of dimensional recurrence equations for Feynman
integrals has been discussed in~\cite{Lee:2009dh}
and implemented
in the code {\tt{SummerTime}}~\cite{Lee:2015eva},
which provides numerical values for the coefficients of the Laurent
series in the $\varepsilon \to 0$ limit, at very high-accuracy
(hundreds of digits). 

Let us observe that $\mathcal{M}_{2,2}$ is finite in
three dimensions, and, within the amplitudes' evaluation, it always
appears multiplied by positive powers of $\varepsilon$, therefore it
drops out of the final result.

In Appendix~\ref{app:masters}, we provide the list of the results for
the MIs of fig.\ref{fig:masters}.  \\ \\

\noindent
{\bf Example.}
As an illustrative example, we apply our algorithm to diagram 49 of
fig.~\ref{diaG5}. 
The corresponding amplitude reads

\begin{eqnarray}
\label{amp49}
\mathcal{A}_{49}&=&
\begin{minipage}{2.5cm}
\begin{center}
\includegraphics[width=2.5cm]{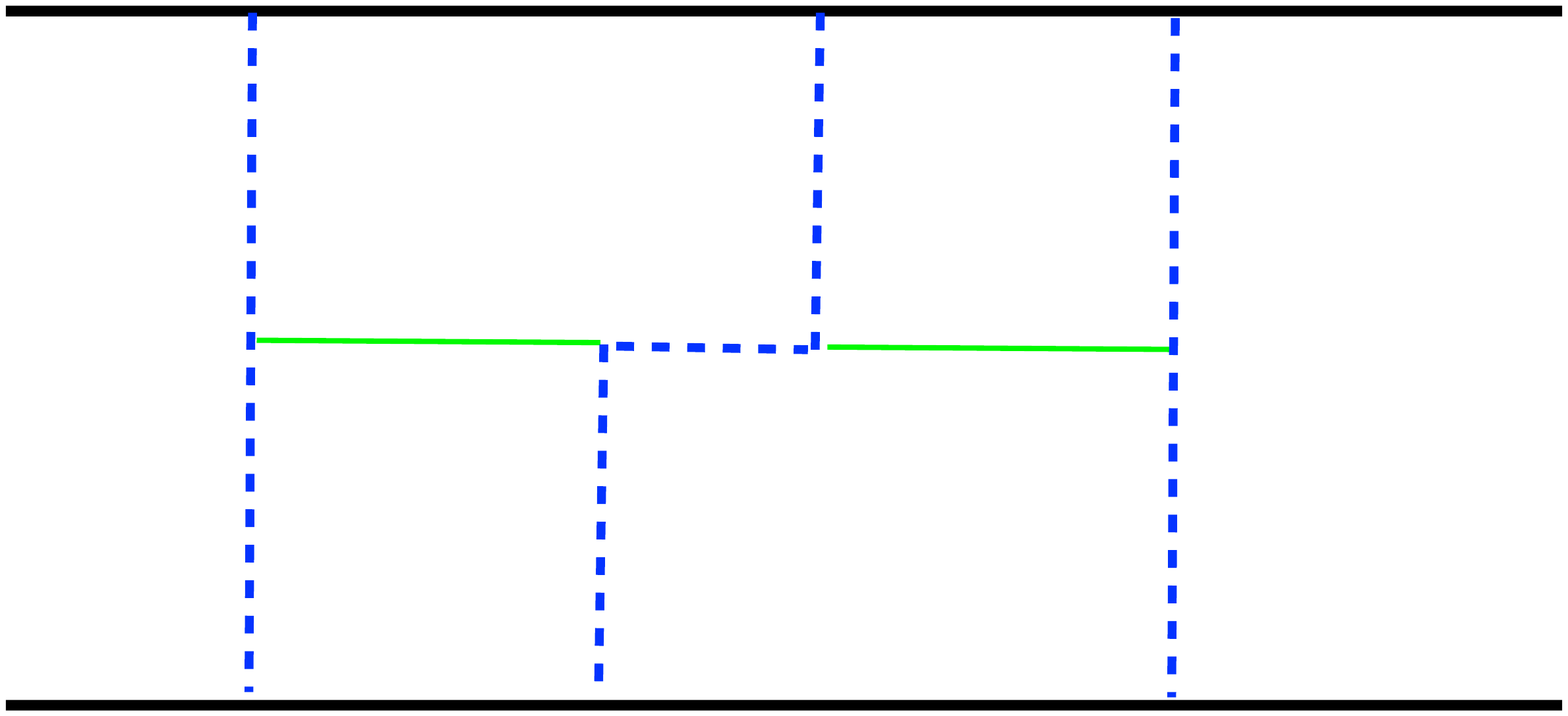}
\end{center}
\end{minipage}
=
-2\ \text{i}\left(8 \pi G_N \right)^5 \left({(d-2) \over (d-1)} \ m_1 m_2\right)^3
\begin{minipage}{2.0cm}
\begin{center}
\includegraphics[width=1.8cm]{topologies/amp49}
\end{center}
\end{minipage} 
\vspace{-1cm}[N_{49}] \ , \qquad
\end{eqnarray}
with
\begin{eqnarray}
\begin{minipage}{2.5cm}
\begin{center}
\includegraphics[width=2.3cm]{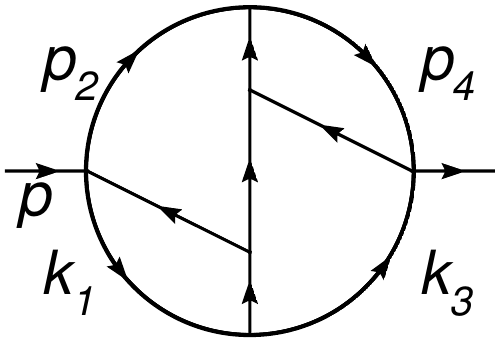}
\end{center}
\end{minipage} 
\vspace{-1cm}[N_{49}] 
&\equiv&
\quad\int_{k_1,k_2,k_3,k_4} \frac{N_{49}}{k_1^2 \ p_2^2 \ k_3^2 \
  p_4^2 \ k_{12}^2 \ k_{13}^2\ k_{23}^2 \ k_{24}^2 \ k_{34}^2} \ , 
\end{eqnarray}
and
\begin{eqnarray}
N_{49} &\equiv&
\left(k_1 \cdot k_3 \ k_{12} \cdot k_{23}-k_1 \cdot k_{12} \ 
k_3 \cdot k_{23} - 
k_1 \cdot k_{23} \ k_3 \cdot k_{12}\right) \times \nonumber \\
& & \qquad
\left(p_2 \cdot k_{23} \ p_4 \cdot k_{34} + p_4 \cdot k_{23} \ 
p_2 \cdot k_{34}-p_2 \cdot p_4 \ k_{23} \cdot k_{34}\right) \ ,
\end{eqnarray}
where we define $\int_k\equiv\int\frac{{\rm d}^d k}{(2\pi)^d}$ and $p_a\equiv p-k_a$, $k_{ab}\equiv k_a-k_b$.
By means of IBPs, we express the 2-point amplitude in terms of MIs,
\begin{eqnarray}
\begin{minipage}{2.0cm}
\begin{center}
\includegraphics[width=1.8cm]{topologies/amp49}
\end{center}
\end{minipage} 
\vspace{-1cm}[N_{49}] 
&=&
c_1
\begin{minipage}{2.0cm}
\begin{center}
\includegraphics[width=1.9cm]{10407014}
\end{center}
\end{minipage}
+
c_2
\begin{minipage}{2.0cm}
\begin{center}
\includegraphics[width=1.9cm]{1040603}
\end{center}
\end{minipage}
+
c_3
\begin{minipage}{2.0cm}
\begin{center}
\includegraphics[width=1.9cm]{1040607}
\end{center}
\end{minipage}
+
\nonumber \\
&+& 
c_4
\begin{minipage}{2.0cm}
\begin{center}
\includegraphics[width=1.9cm]{1040602}
\end{center}
\end{minipage}
+
c_5
\begin{minipage}{2.0cm}
\begin{center}
\includegraphics[width=1.9cm]{1040501}
\end{center}
\end{minipage}
\ ,
\end{eqnarray}
with
\begin{eqnarray} 
c_1&=&{(d-3)^2\*(d-2)^2\*s^2\over(d-4)^2\*(5\*d-14)\*(12-5\*d)},\quad
c_2 = {(d-2)^2\*(432-512\*d+203\*d^2-27\*d^3)\*s\over8\*(d-4)^3\*(5-2\*d)\*(5\*d-12)},\;\;\\
c_3&=&{(d-2)^2\*(76-58\*d+11\*d^2)\*s\over4\*(d-4)^2\*(14-5\*d)\*(5\*d-12)},\quad
c_4 = {(d-2)^2\*s\over2\*(d-4)^2},\\
c_5&=&{(d-2)^2\*(1096-1598\*d+870\*d^2-210\*d^3+19\*d^4)\over(d-4)^4\*(3-d)\*(3\*d-8)}.
\end{eqnarray} 
This result can be expanded around $d=3+\varepsilon$, using the
expressions of the MIs given in Appendix A,
\begin{eqnarray}
\mathcal{A}_{49}&=&-\text{i}
(8\*\pi\*G_N)^5\*(m_1\*m_2)^3\*2^{-4}\*(4\*\pi)^{-(4+2\*\vep)}\*e^{2\*\vep\*\gamma_E}\* s^{(1+2\*\vep)}
\times 
\nonumber\\&&
\qquad \left[
 {1\over\vep}\*\left({\pi^2\over16}-{2\over3}\right)
+ {29\over18} 
- {13\over144}\*\pi^2 
- {\pi^2\over8}\*\log{2}
+ \mathcal{O}(\vep)
\right],
\end{eqnarray}
where $\gamma_E=0.57721...$ is the Euler-Mascheroni constant.
Finally, by means of the Fourier transform formula
\be\label{Fourier}
\int_p {\rm e}^{\text{i}p \cdot r} p^{-2 a}= \frac{\Gamma(d/2-a)}{(4\pi)^{d/2}\Gamma(a)}
\left(\frac{r}{2}\right)^{(2 a -d)}\,,
\ee
one obtains the following Lagrangian term,
\be\label{eq:res:A49}
\mathcal{L}_{49} \ = - \text{i} \lim_{d \to 3} \ 
\int_p {\rm e}^{\text{i}p \cdot r} \mathcal{A}_{49}
=(32- 3 \pi^2)\frac{G_N^5 m_1^3 m_2^3}{r^5}\,.
\ee

\section{Results and discussion}
\label{sec:discussion}
The complete 4PN, ${\cal O}(G_N^5)$ Lagrangian was already presented in \cite{Bernard:2015njp},
\begin{eqnarray}\label{blanchet}
{\cal L}_{4PN}^{G_N^5} &=& 
\frac{3}{8}\frac{G_N^5 m_1^5 m_2}{r^5}+\frac{G_N^5 m_1^4 m_2^2}{r^5}\left[\frac{1690841}{25200}+\frac{105}{32}\pi^2-\frac{242}{3}\log\frac{r}{r'_1}-16\log\frac{r}{r'_2}\right]\nonumber\\
&+&\frac{G_N^5 m_1^3 m_2^3}{r^5}\left[\frac{587963}{5600}-\frac{71}{32}\pi^2-\frac{110}{3}\log\frac{r}{r'_1}\right]+
(m_1\leftrightarrow m_2) \ ,
\end{eqnarray}
where $r'_1, r'_2$ are two UV scales which do not contribute to physical observables.
Such a Lagrangian gets contributions from the 50 genuine ${\cal O}(G_N^5)$ diagrams depicted in fig.\ref{diaG5},
and from diagrams at lower orders in $G_N$ which are at least quadratic in the accelerations:
\begin{eqnarray}
{\cal L}_{4PN}^{G_N^5} = \sum_{a=1}^{50} {\cal L}_a  + \sum_{j=1}^3{\cal L}_{4PN}^{G_N^j\rightarrow G_N^5} +
(m_1\leftrightarrow m_2)\ . 
\label{sym12}
\end{eqnarray}
The evaluation of  $\sum_{a=1}^{50} {\cal L}_a $ represents the main result of this work, and it amounts to
\begin{eqnarray}
\sum_{a=1}^{50} {\cal L}_a &=&
\frac{3}{8}\frac{G_N^5 m_1^5 m_2}{r^5}+\frac{31}{3}\frac{G_N^5 m_1^4 m_2^2}{r^5}
+\frac{141}{8}\frac{G_N^5 m_1^3 m_2^3}{r^5}
\label{ourresult}\,.
\end{eqnarray}

The individual contributions ${\cal L}_a$ are presented in Appendix B. 
We observe that, although there appear contributions which are divergent in the $d \to 3$ limit, the sum of all contributions is finite,
hence $L$ does not show up in physical observables.

To obtain the whole expression for the 4PN ${\cal O}(G_N^5)$ corrections, one would need to add contributions generated from lower $G_N$ terms when using the equations of motion, in order to eliminate terms quadratic at least in the accelerations.
All such contributions have been computed also in the EFT framework \cite{Foffa:2012rn}, except for ${\cal L}_{4PN}^{G_N^3\rightarrow G_N^5}$.
We can nevertheless perform partial checks between eq.(\ref{ourresult}) and eq.(\ref{blanchet}). \\

\noindent
{\bf The $m_1^5 m_2$-term}. It can be proven that this term does not
receive any contribution from lower $G_N$ terms\footnote{
Contributions to this term from lower $G_N$ orders would come from terms
of the type $G_N^{5-n}m_1^{5-n}m_2 a_2^n$ with $2\leq n\leq 4$. However, diagrams
giving rise to such terms would have exactly one propagator attached to particle
 2, hence $a_2^2$ or higher power of $a_2$ can be taken out by integration by parts
instead of by using the doube zero trick. It can be checked explicitly in
\cite{Foffa:2012rn} that $G_N^{5-n}m_1^{5-n}m_2 a_2^n$ terms do not appear in the
Lagrangian for $n=3,4$.},
and the corresponding coefficient for the two-body Lagrangian of
eq.(\ref{ourresult}) agrees with the Lagrangian term reported in
eq.(\ref{blanchet}).\\ 

\noindent
{\bf The $\pi^2$-term.} The contributions coming from the lower $G_N$ orders come entirely from the still unpublished ${\cal L}_{4PN}^{G_N^3\rightarrow G_N^5}$: for dimensional reasons terms at least quadratic in the accelerations can appear only in $G_N^{m\leq n-1}$ sectors at $n$-th PN order, and all the terms up to ${\cal O}(G_N^2)$ do not contain $\pi^2$.
Although the computational details will be given elsewhere, such contributions have been computed in the EFT framework and found to be 
\be\label{4PNG3}
\frac{105}{32}\pi^2\frac{G_N^5 m_1^4 m_2^2}{r^5}-\frac{71}{32}\pi^2\frac{G_N^5 m_1^3 m_2^3}{r^5}\ . 
\ee
This result, alone, already accounts for the Lagrangian $\pi^2$-term of eq.~(\ref{blanchet}), presented in \cite{Bernard:2015njp} and previously computed also in \cite{Damour:2015isa}. 
Athough some of the ${\cal L}_a$'s listed in Appendix B (namely, $a=33,49,50$) contain terms proportional to $\pi^2$, these terms cancel in the sum of all the diagrams (as shown in ref.~\cite{Damour:2017ced}), thus providing agreement with the literature.

\noindent
{\bf Other terms.} The other terms are not directly comparable without full knowledge of the ${\cal L}_{4PN}^{G_N^3\rightarrow G_N^5}$ contribution, and without taking into account the different regularization schemes used here and in \cite{Bernard:2015njp}.

\section{Conclusion}
\label{sec:conclusion}
We studied the conservative dynamics of the two-body motion at fourth
post-Newtonian order (4PN), at fifth order in the Newton constant $G_N$,
within the effective field theory (EFT) framework to General Relativity.
We determined an essential contribution of the complete 4PN Lagrangian
at ${\cal O}(G_N^5)$, coming from 50 Feynman diagrams.  By exploiting
the analogy between such diagrams in the EFT gravitational theory and
2-point 4-loop functions in massless gauge theory, we addressed their
calculation by means of multi-loop diagrammatic techniques, based on
integration-by-parts identities and difference equations.  We performed
the calculation within the dimensional regularization scheme, and the
contribution to the Lagrangian of each graph was given as Laurent series
in $d=3 + \varepsilon$, being $d$ the number of dimensions. Although
some individual amplitudes are divergent in the $\vep \to 0$ limit and others contain the irrational factor $\pi^2$, the
sum of the fifty terms is found to be finite at $d=3$ and rational, in agreement with previous calculations
performed with other techniques.\\

\section*{Notes}

In a first version of this manuscript, $\mathcal{L}_{50}$ appeared to
have a different value, yielding to a disagreement with the literature.
Subsequently, the authors of ref.~\cite{Damour:2017ced} pointed us to a missing
overall factor of ``$-3$'' in $\mathcal{L}_{50}$,
which we have been able to find and correct: 
the value of $\mathcal{L}_{50}$ reported in this version is the amended one.
Let us also notice, that the analytic result for the master integral
${\cal M}_{3,6}$ obtained
in \cite{Damour:2017ced} agrees with the semi-analytic expression given in our current work. 

\acknowledgments

We thank Luc Blanchet, Thibault Damour, Guillaume Faye and
Ulrich Schubert-Mielnik for clarifying discussions, and Andreas von 
Manteuffel for kind correspondence on the use of {\tt Reduze}.
We wish to thank ICTP-SAIFR, supported by FAPESP grant 2016/01343-7,
for the organization of the workshop ``Analytic methods in General Relativity",
where many stimulating discussions took place.
The work of RS has been supported for most of the duration of the present work
by the FAPESP grant n. 2012/14132-3 and by the High Performance Computing Center at UFRN.
SF is supported by the Fonds National Suisse and by the SwissMap NCCR.


\appendix
\section{Master integrals\label{app:masters}}
In this appendix, we provide the expressions of the master
integrals. They are defined by
\begin{align*}
 \mathcal{M}_{0,1}&=\int_{k_{1...4}}{1\over D_{1...4} D_{14}},
&\mathcal{M}_{1,1}&=\int_{k_{1...4}}{1\over D_{1...4} D_{9} D_{12}},\\
\mathcal{M}_{1,2}&=\int_{k_{1...4}}{1\over D_{1...4} D_{10} D_{11}},
&\mathcal{M}_{1,3}&=\int_{k_{1...4}}{1\over D_{1...4} D_{8} D_{10}},\\
\mathcal{M}_{1,4}&=\int_{k_{1...4}}{1\over D_{1...4} D_{7} D_{13}},
&\mathcal{M}_{2,2}&=\int_{k_{1...4}}{1\over D_{1...4} D_{10} D_{15} D_{16}},\\
\mathcal{M}_{3,6}&=\int_{k_{1...4}}{1\over D_{1...4} D_{5} D_{6} D_{10} D_{14}},
\end{align*}
where $k_i$ ($i=1,2,3,4$) are the loop momenta and $p$ is the external
momentum of the diagrams depicted in fig.~\ref{fig:masters}. The
integral measure is the same as used in sec.~\ref{sec:results} and given by
$\int_{k_{1...4}}=\int_{k_{1}}\int_{k_{2}}\int_{k_{3}}\int_{k_{4}}$ with
$\int_{k_{i}}\equiv\int\frac{{\rm d}^d k_i}{(2\pi)^d}$ ($i=1,2,3,4$).  
The denominators read
\begin{align*}
 D_{1...4}&=k_1^2\,k_2^2\,k_3^2\,k_4^2,
&D_{5 }&=(k_2-k_3)^2,
&D_{6 }&=(k_1-k_4)^2,&&\\
 D_{7 }&=(k_2+k_3-k_4)^2,
&D_{8 }&=(k_1+k_2+k_3-k_4)^2,
&D_{9 }&=(k_1-p)^2,&&\\
 D_{10}&=(k_1+k_2-p)^2,
&D_{11}&=(k_3+k_4+p)^2,
&D_{12}&=(k_2-k_3-k_4+p)^2,&&\\
D_{13}&=(k_1-k_2-k_3+p)^2,
&\omit\rlap{$D_{14}=(k_1+k_2-k_3-k_4-p)^2$,}\\
 D_{15}&=(k_1+k_4-p)^2,
&D_{16}&=(k_2+k_3-p)^2.
\end{align*}
\subsection{Master integrals known in $d$ dimensions}
The following master integrals are known in closed analytical form,
exact in $d$: 
\begin{eqnarray}
\label{eq:M01}
\mathcal{M}_{0,1}&=&(4\*\pi)^{-2\*d}\*s^{2\*d-5}\*
{\Gamma(5-2\*d)\*\Gamma({d\over2}-1)^5\over\Gamma\!\left({5\over2}\*d-5\right)}\\
&\stackrel{d=3+\vep}{=}&c(\vep)\*s\*\left[
  {1 \over 24\*\vep}
- {13\over36} 
+ \vep\*\left(
    {481\over 216}
  - {11\over 288}\*\pi^2
        \right) 
\right.\nonumber\\&&\left.
- \vep^2\*\left(
  {3943\over 324}
- {143\over 432}\*\pi^2 
- {113\over 72}\*\z3
          \right) 
+\mathcal{O}(\vep^3)\right],\\
\mathcal{M}_{1,1}&=&(4\*\pi)^{-2\*d}\*s^{2\*d-6}\*
{\Gamma\!\left(4-{3\over2}\*d\right)\*\Gamma\!\left(2-{d\over2}\right)\*\Gamma\!\left({d\over2}-1\right)^6\over\Gamma(d-2)\*\Gamma(2\*d-4)}\\
&\stackrel{d=3+\vep}{=}&
-c(\vep)\*\pi^2\*\left[
 {1\over8}
+\mathcal{O}(\vep^1)
\right],\\
\mathcal{M}_{1,2}&=&(4\*\pi)^{-2\*d}\*s^{2\*d-6}\*
{\Gamma(3-d)^2\*\Gamma\!\left({d\over2}-1\right)^6\over\Gamma\!\left({3\over2}\*d-3\right)^2}\\
&\stackrel{d=3+\vep}{=}&
c(\vep)\*\left[
   {1\over4\*\vep^2} 
- {3\over2\*\vep} 
+ \left({27\over4} - {7\over48}\*\pi^2\right)
\right.\nonumber\\&&\left.
- \vep\*\left(27 - {7\over8}\*\pi^2 - {11\over3}\*\z3 \right)
+\mathcal{O}(\vep^2)\right],\\
\mathcal{M}_{1,3}&=&(4\*\pi)^{-2\*d}\*s^{2\*d-6}\*
{\Gamma(6-2\*d)\*\Gamma(3-d)\*\Gamma\!\left(2-{d\over2}\right)\*\Gamma\!\left({d\over2}-1\right)^6\*\Gamma(2\*d-5)\over
\Gamma\!\left(5-{3\over2}\*d\right)\*\Gamma(d-2)\*\Gamma\!\left({3\over2}\*d-3\right)\*\Gamma\!\left({5\over2}\*d-6\right)}\\
&\stackrel{d=3+\vep}{=}&
c(\vep)\*\left[
  {1\over8\*\vep^2} 
- {1\over\vep} 
+ {49\over8} - {19\over96}\*\pi^2
\right.\nonumber\\&&\left.
- \vep\*\left( 34 - {19\over12}\*\pi^2 - {107\over24}\*\z3\right)
+ \mathcal{O}(\vep^2)\right],\\
\mathcal{M}_{1,4}&=&(4\*\pi)^{-2\*d}\*s^{2\*d-6}\*
{\Gamma(6-2\*d)\*\Gamma\!\left(2-{d\over2}\right)^2\*\Gamma\!\left({d\over2}-1\right)^6\*\Gamma\!\left({3\over2}\*d-4\right)\over\Gamma(4-d)\*\Gamma(d-2)^2\*\Gamma\!\left({5\over2}\*d-6\right)}\\
&\stackrel{d=3+\vep}{=}&
-c(\vep)\*\pi^2\*\left[
  {1\over16\*\vep}
- \left({5\over16} + {1\over8}\*\log{2}\right) 
+ \mathcal{O}(\vep^1)\right],
\label{eq:M14exp}
\end{eqnarray}
with the Euler $\Gamma$ function
$\Gamma(z)=\int_0^{\infty}t^{z-1}e^{-t} {\rm d}t$, 
the Riemann zeta function $\zeta_n=\sum_{k=1}^{\infty}{1\over k^n}$,
and $s=p^2$. The coefficient function $c(\vep)$ is given by
\begin{equation}
c(\vep)=e^{2\*\vep\*\gamma_E}\*s^{2\*\vep}/(4\*\pi)^{4+2\*\vep}.
\end{equation}
\subsection{Master integrals known in $d = 3 + \vep$ dimensions}
The master integrals $\mathcal{M}_{2,2}$ and $\mathcal{M}_{3,6}$ are
known numerically~\cite{Lee:2015eva}. In three dimensions
$\mathcal{M}_{2,2}$ is finite,
i.e. $\mathcal{M}_{2,2}=\mathcal{O}\left(\vep^0\right)$, and does not
contribute to our amplitudes, since it always appears multiplied by a
positive power of $\vep$. The Laurent expansion in $\vep$ around $d=3$
for $\mathcal{M}_{3,6}$ reads,
\begin{eqnarray}
\mathcal{M}^{d=3+\vep}_{3,6}&=&
{c(\vep)\over s^{2}}\*\left[
\right.\nonumber\\&&\left.\phantom{+}
 0.50000000000000000000000000000000000000000000000000000000000/\vep^2
\nonumber\right.\\&&\left.
-0.50000000000000000000000000000000000000000000000000000000000/\vep
\nonumber\right.\\&&\left.
-3.58876648328794339088189620833849370269526252469830039056611
\nonumber\right.\\&&\left.
+15.6234156117945512067218751269082577384023065736147735689317\,\vep
\nonumber\right.\\&&\left.
+\mathcal{O}\left(\vep^2\right)
\right]\\
&\stackrel{\mbox{\tiny{PSLQ}}}{\hateq}&
{c(\vep)\over s^{2}}\*\left[
 {1\over2\*\vep^2}
-{1\over2\*\vep}
-4+{\pi^2\over24}
-\vep
\*\left(
9 - \pi^2\*\left({13\over8} - \log{2}\right) - {77\over6}\*\z3
\right)
+\mathcal{O}\left(\vep^2\right)
\right].
\label{M36}
\end{eqnarray}
The analytical coefficients in the $\vep$ expansion have been obtained
from the high precision numerical result with the PSLQ algorithm
\cite{PSLQ}. 
We observe that, according to the arguments in footnote 4, the value
of the coefficient of the double pole can be obtained analytically
from the recurrence relation: its numerical reconstruction agrees with
the analytic determined value.

Moreover, in order to perform a consistency check of the other analytical coefficients
of eq.~(\ref{M36}), we determined $\mathcal{M}_{3,6}$ also in 1- and
5-dimensions with {\tt{SummerTime}}~\cite{Lee:2015eva} numerically
and used the PSLQ algorithm to obtain again the analytical
coefficients of the $\vep$ expansion, respectively reading,
\begin{eqnarray}
\mathcal{M}^{d=1+\vep}_{3,6}&=&
(4\*\pi)^4\*{c(\vep)\over s^{6}}\*\left[
\right.\nonumber\\&&\left.\phantom{+}
 11.0000000000000000000000000000000000000000000000000000000000/\vep
\nonumber\right.\\&&\left.
+750.157936507936507936507936507936507936507936507936507936508
\nonumber\right.\\&&\left.
-5333.19383013044510985261411265298578814107960018433010670281\,\vep
\nonumber\right.\\&&\left.
-3509.80936167055655677303026105319710926833682220819489993426\,\vep^2
\nonumber\right.\\&&\left.
+\mathcal{O}\left(\vep^3\right)
\right]\\
&\stackrel{\mbox{\tiny{PSLQ}}}{\hateq}&
(4\*\pi)^4\*{c(\vep)\over s^{6}}\*\left[
  {11\over\vep}
+ {945199\over1260}
- \vep\*\left(
  {35338924\over6615}
- {11\over12}\*\pi^2
         \right) 
+ \vep^2\*\left(
  {160485605363\over27783000 }
\nonumber\right.\right.\\&&\left.\left.
- {14515601\over15120}\*\pi^2 
- 22\*\pi^2\*\log{2} + 
  {847\over3}\*\z3
  \right)
+\mathcal{O}\left(\vep^3\right)
\right],
\label{M36d1}
\end{eqnarray}
\begin{eqnarray}
\mathcal{M}^{d=5+\vep}_{3,6}&=&
{1\over (4\*\pi)^4}\*{c(\vep)\*s^{2}\over 2520}\*\left[
\right.\nonumber\\&&\left.\phantom{+}
 1.00000000000000000000000000000000000000000000000000000000000/\vep^2
\nonumber\right.\\&&\left.
-7.49665930774956257270733971502880747383208927084097052723419/\vep
\nonumber\right.\\&&\left.
+33.1813244635562837450781924787207309198665172698916969562612
\nonumber\right.\\&&\left.
+\mathcal{O}\left(\vep\right)
\right]\\
&\stackrel{\mbox{\tiny{PSLQ}}}{\hateq}&
{1\over (4\*\pi)^4}\*{c(\vep)\*s^{2}\over 2520}\*\left[
  {1\over\vep^2}
- {1\over\vep}\*\left({467\over7} - 6\*\pi^2\right)
\nonumber\right.\\&&\left.
+ {123478\over147} - {1651\over 21}\*\pi^2 
+ 54\*\pi^2\*\log{2} - 333\*\z3
+\mathcal{O}\left(\vep\right)
\right].
\label{M36d5}
\end{eqnarray}
We verified that the analytical
ans{\"a}tze for $\mathcal{M}^{d=1+\vep}_{3,6}$,
$\mathcal{M}^{d=3+\vep}_{3,6}$, 
$\mathcal{M}^{d=5+\vep}_{3,6}$
fulfill the dimensional recurrence
relation~(\ref{eq:recrel:M36}) analytically, order-by-order in $\vep$,
therefore we have high
confidence in their correctness.

\section{Results for all the amplitudes}
\label{app:50amplitudes}
In this appendix we collect the contributions to the Lagrangian in eq.~(\ref{sym12}), coming from all the amplitudes of
fig.~\ref{diaG5}:
\begin{equation}
0=\mathcal{L}_9=\mathcal{L}_{12}=\mathcal{L}_{13}=\mathcal{L}_{22}=\mathcal{L}_{26}
=\mathcal{L}_{27}=\mathcal{L}_{31}=\mathcal{L}_{36}=\mathcal{L}_{46}=\mathcal{L}_{47}\,,\nonumber
\end{equation}
\begin{equation}
\frac{1}{2}\frac{G_N^5 m_1^3 m_2^3}{r^5}=\mathcal{L}_1=\mathcal{L}_3=4\mathcal{L}_5=3\mathcal{L}_{14}=\frac{\mathcal{L}_{19}}{8}=\frac{3\mathcal{L}_{20}}{2}=\frac{3\mathcal{L}_{21}}{4}=\frac{\mathcal{L}_{23}}{4}=\frac{\mathcal{L}_{24}}{4}=\frac{3\mathcal{L}_{25}}{2}\,,\nonumber
\end{equation}
\begin{equation}
\frac{1}{2}\frac{G_N^5 m_1^4 m_2^2}{r^5}=\mathcal{L}_2=3\mathcal{L}_4=\frac{3\mathcal{L}_8}{2}=\frac{3\mathcal{L}_{10}}{2}=\frac{3\mathcal{L}_{11}}{2}=\frac{\mathcal{L}_{15}}{4}=\frac{3\mathcal{L}_{16}}{4}=\frac{3\mathcal{L}_{17}}{4}=\frac{\mathcal{L}_{18}}{4}\,,\nonumber
\end{equation}
\begin{equation}
\frac{1}{120}\frac{G_N^5 m_1^5 m_2}{r^5}=\mathcal{L}_6=\frac{\mathcal{L}_{7}}{20}=\frac{3\mathcal{L}_{30}}{20}=-\frac{3\mathcal{L}_{35}}{56}=\frac{\mathcal{L}_{39}}{24}=\frac{\mathcal{L}_{45}}{12}\,,\nonumber
\end{equation}
\begin{eqnarray}
\mathcal{L}_{28}=\frac{G_N^5 m_1^4 m_2^2}{r^5}\left[\frac{428}{75}+\frac{4}{15}{\cal P}\right]\,,\quad\quad\quad\quad\quad\quad&&\mathcal{L}_{29}=\frac{G_N^5 m_1^3 m_2^3}{r^5}\left[-\frac{409}{450}+\frac{1}{5}{\cal P}\right]\,,\nonumber\\
\mathcal{L}_{32}=\frac{G_N^5 m_1^3 m_2^3}{r^5}\left[-\frac{91}{450}+\frac{1}{15}{\cal P}\right]\,,\quad\quad\quad\quad\quad\quad&&\mathcal{L}_{33}=\frac{G_N^5 m_1^3 m_2^3}{r^5}\left(16-\pi^2\right)\,,\nonumber\\
\mathcal{L}_{34}=\frac{G_N^5 m_1^4 m_2^2}{r^5}\left[\frac{13}{5}-\frac{2}{3}{\cal P}\right]\,,\quad\quad\quad\quad\quad\quad&&\mathcal{L}_{37}=-\frac{G_N^5 m_1^4 m_2^2}{r^5}\left[17+2{\cal P}\right]\,,\nonumber\\
\mathcal{L}_{38}=\frac{G_N^5 m_1^4 m_2^2}{r^5}\left[\frac{147}{25}+\frac{8}{15}{\cal P}\right]\,,\quad\quad\quad\quad\quad\quad&&\mathcal{L}_{40}=\frac{G_N^5 m_1^4 m_2^2}{r^5}\left[-\frac{39}{25}+\frac{4}{15}{\cal P}\right]\,,\nonumber\\
\mathcal{L}_{41}=\frac{G_N^5 m_1^3 m_2^3}{r^5}\left[\frac{49}{18}+\frac{1}{3}{\cal P}\right]\,,\quad\quad\quad\quad\quad\quad&&\mathcal{L}_{42}=-\frac{G_N^5 m_1^3 m_2^3}{r^5}\left[\frac{97}{225}+\frac{1}{15}{\cal P}\right]\,,\nonumber\\
\mathcal{L}_{43}=-\frac{G_N^5 m_1^3 m_2^3}{r^5}\left[\frac{53}{150}+\frac{2}{15}{\cal P}\right]\,,\quad\quad\quad\quad\quad\quad&&\mathcal{L}_{44}=-\frac{G_N^5 m_1^3 m_2^3}{r^5}\left[\frac{37}{75}+\frac{2}{5}{\cal P}\right]\,,\nonumber\\
\mathcal{L}_{48}=\frac{G_N^5 m_1^4 m_2^2}{r^5}\left[\frac{578}{75}+\frac{8}{5}{\cal P}\right]\,,\quad\quad\quad\quad\quad\quad&&\mathcal{L}_{49}=\frac{G_N^5 m_1^3 m_2^3}{r^5}\left(32-3\pi^2\right)\,,\nonumber\\
\mathcal{L}_{50}=\frac{G_N^5 m_1^3 m_2^3}{r^5}\left(4 \pi^2-\frac{124}{3}\right)\,,\quad\quad\quad\quad\quad\quad&&
\end{eqnarray}
where the pole part ${\cal P}\equiv\frac{1}{\vep}-5 \log{\frac{r}{L_0}}$ (with $L_0$ defined by  $L=\sqrt{4 \pi{\rm e}^{\gamma_E}} L_0$) cancels exactly in the sum of all the terms.

Diagrams which are symmetric under $(1\leftrightarrow 2)$ exchange, i.e. 3, 5, 22, 23, 24, 32, 33, 41, 42, 43, 49, 50 have been multiplied by $1/2$.

\section{Evaluation of $\mathcal{A}_{33}$ and $\mathcal{A}_{50}$}
\label{app:a33_a50}
We describe the evaluation of amplitudes 33 and 50 which, along with amplitude 49 already discussed in
detail in section \ref{sec:results}, are the only ones containing $\pi^2$ terms.

\subsection{Amplitude 33}
\begin{eqnarray}
\label{amp33}
\mathcal{A}_{33}&=&
\begin{minipage}{2.5cm}
\begin{center}
\includegraphics[width=2.5cm]{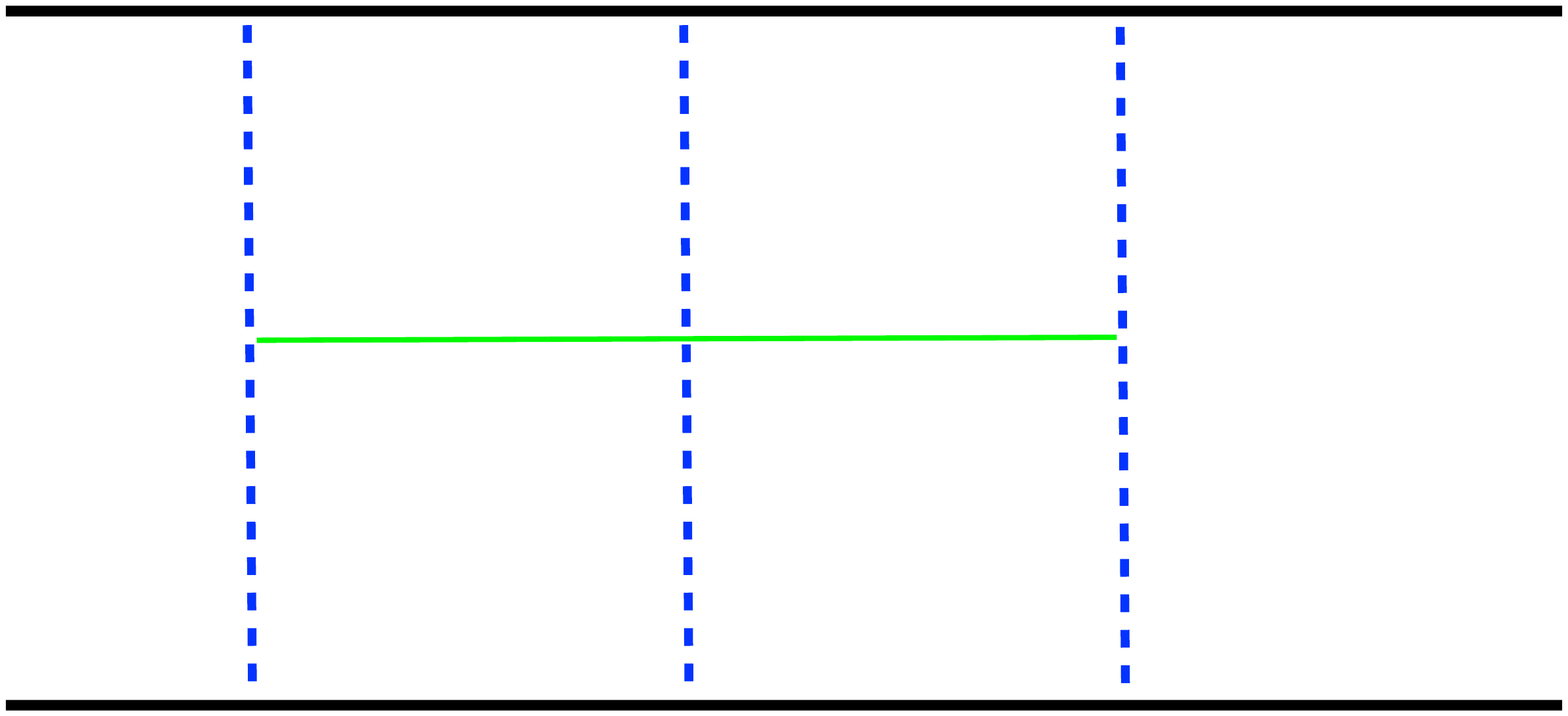}
\end{center}
\end{minipage}
=
-\ \text{i}\left(8 \pi G_N \right)^5 \left({(d-2) \over (d-1)} \ m_1 m_2\right)^3
\begin{minipage}{2.0cm}
\begin{center}
\includegraphics[width=1.8cm]{topologies/amp33}
\end{center}
\end{minipage} 
\vspace{-1cm}[N_{33}] \ , \qquad
\end{eqnarray}
with
\begin{eqnarray}
\begin{minipage}{2.0cm}
\begin{center}
\includegraphics[width=1.8cm]{topologies/amp33}
\end{center}
\end{minipage} 
\vspace{-1cm}[N_{33}] 
&\equiv&
\quad\int_{k_1,k_2,k_3,k_4} \frac{N_{33}}{k_1^2 \ k_2^2 \ k_3^2 \
  k_4^2 \ k_{14}^2 \ p_{12}^2\ p_{34}^2 \ p_{123}^2} \ , 
\end{eqnarray}
and
\begin{eqnarray}
N_{33} &\equiv&
k_3\cdot k_4 \left(k_2\cdot p_{12}\ k_1\cdot p_{34} + k_1\cdot k_2\ p_{12}\cdot p_{34} - k_1\cdot\ p_{12} \ k_2\cdot p_{34}\right) \nonumber \\
& &+ k_2\cdot k_4 \left(k_1\cdot\ p_{12} \ k_3\cdot p_{34}+ 
k_1\cdot k_3\ p_{12}\cdot p_{34} -k_3\cdot p_{12}\ k_1\cdot p_{34}\right)\nonumber \\
& & +k_1\cdot k_4 \left(k_3\cdot\ p_{12} \ k_2\cdot p_{34}-k_2 \cdot p_{12}\ k_3\cdot p_{34} -k_2\cdot k_3\ p_{12}\cdot p_{34} \right)\nonumber \\
& & +k_2\cdot k_3 \left(k_4\cdot p_{12}\ k_1\cdot p_{34} + k_1\cdot\
  p_{12}\ k_4\cdot p_{34}\right)\nonumber\\
& &
+k_1\cdot k_3 \left(k_2\cdot\ p_{12}\ k_4\cdot p_{34}-k_4\cdot p_{12}\ k_2 \cdot p_{34} \right)\nonumber\\
& &+k_1\cdot k_2 \left(k_4\cdot \ p_{12}\ k_3\cdot p_{34}-k_3\cdot p_{12}\ k_4\cdot p_{34}\right)\,,
\end{eqnarray}
where $p_{123}\equiv p-k_1-k_2-k_3$, $p_{ab}\equiv p-k_a-k_b$, $k_{14}\equiv k_1-k_4$. By means of IBPs, we express the 2-point amplitude in terms of MIs,
\begin{eqnarray}
\begin{minipage}{2.0cm}
\begin{center}
\includegraphics[width=1.8cm]{topologies/amp33}
\end{center}
\end{minipage} 
\vspace{-1cm}[N_{33}] 
&=&
c_1
\begin{minipage}{2.0cm}
\begin{center}
\includegraphics[width=1.9cm]{1040802}
\end{center}
\end{minipage}
+
c_2
\begin{minipage}{2.0cm}
\begin{center}
\includegraphics[width=1.9cm]{1040603}
\end{center}
\end{minipage}
+
c_3
\begin{minipage}{2.0cm}
\begin{center}
\includegraphics[width=1.9cm]{1040607}
\end{center}
\end{minipage}
+
\nonumber \\
&+& 
c_4
\begin{minipage}{2.0cm}
\begin{center}
\includegraphics[width=1.9cm]{1040602}
\end{center}
\end{minipage}
+
c_5
\begin{minipage}{2.0cm}
\begin{center}
\includegraphics[width=1.9cm]{1040501}
\end{center}
\end{minipage}
\end{eqnarray}
and
\begin{eqnarray} 
c_1&=&{(d-2)\*(3\*d-10)\*(d^2-12\*d+24)\*s^3\over4\*(d-3)\*(5\*d-16)\*(5\*d-14)\*(5\*d-12)},\\ 
c_2 &=& {(d-2)\*(19\*d^4+225\*d^3-2708\*d^2+8140\*d-7680)\*s\over4\*(d-4)^2\*(2\*d-5)\*(3\*d-10)\*(5\*d-12)},\;\;\\
c_3&=&{(d-2)\*(33\*d^5-44\*d^4-1936\*d^3+11024\*d^2-22512\*d+16128)\*s\over4\*(d-4)^2\*(d-3)\*(5\*d-16)\*(5\*d-14)\*(5\*d-12)},\\
c_4 &=& -{2(d-2)\*(d^3+7\*d^2-55\*d+78)\*s\over (d-4)^2\*(d-3)\*(5\*d-12)},\\
c_5&=&{(d-2)\*(2\*d -5)\*(3\*d^4+204\*d^3-1856\*d^2+5296\*d-4944)\over2\*(d-4)^2\*(d-3)^2\*(3\*d-10)\*(3\*d-8)}.
\end{eqnarray} 
This result can be expanded around $d=3+\varepsilon$, using the
expressions of the MIs given in Appendix A,
\begin{eqnarray}
\mathcal{A}_{33}&=&-\text{i}
(8\*\pi\*G_N)^5\*(m_1\*m_2)^3\*2^{-4}\*(4\*\pi)^{-(4+2\*\vep)}\*e^{2\*\vep\*\gamma_E}\* s^{(1+2\*\vep)}
\times 
\nonumber\\&&
\qquad \left[
 {1\over\vep}\*\left({\pi^2\over48}-{1\over3}\right)
+{49\over 18} 
- {5\pi^2\over 16} 
+ {7\pi^2\over8}\*\log{2}
- {37\zeta_3\over8}
+ \mathcal{O}(\vep)
\right].
\end{eqnarray}
Finally, by applying the Fourier transform formula (\ref{Fourier}) to $-\text{i} {\cal A}_{33}$, one gets the result for ${\cal L}_{33}$ reported in appendix B.

\subsection{Amplitude 50}
Coming to amplitude 50, we have
\begin{eqnarray}
\label{amp50}
\mathcal{A}_{50}&=&
\begin{minipage}{2.5cm}
\begin{center}
\includegraphics[width=2.5cm]{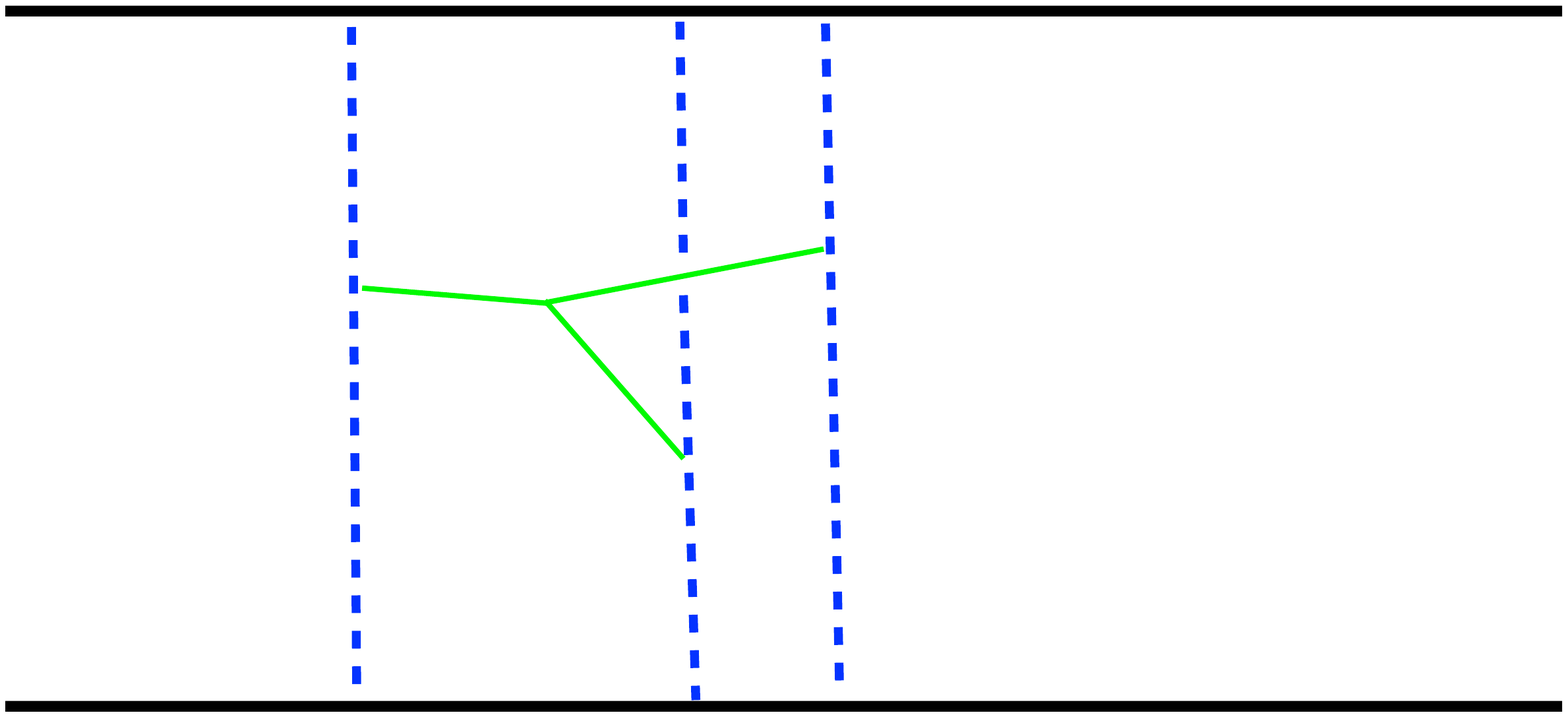}
\end{center}
\end{minipage}
=
- \text{i}\left(8 \pi G_N \right)^5 \left({(d-2) \over (d-1)} \ m_1 m_2\right)^3
\begin{minipage}{2.0cm}
\begin{center}
\includegraphics[width=1.8cm]{topologies/amp50}
\end{center}
\end{minipage} 
\vspace{-1cm}[N_{50}] \ , \qquad
\end{eqnarray}
with
\begin{eqnarray}
\begin{minipage}{2.0cm}
\begin{center}
\includegraphics[width=1.8cm]{topologies/amp50}
\end{center}
\end{minipage} 
\vspace{-1cm}[N_{50}] 
&\equiv&
\quad\int_{k_1,k_2,k_3,k_4} \frac{N_{50}}{k_1^2 \ k_2^2 \ k_3^2 \
  k_4^2 \ k_{12}^2 \ k_{34}^2\ \hat{k}_{24}^2 \ p_{13}^2 \ \hat{p}_{14}^2} \ , 
\end{eqnarray}
and
\begin{eqnarray}
N_{50} &\equiv&
(k_3\cdot p_{13}\  k_{12}\cdot\hat{p}_{14}-k_{12}\cdot p_{13}\
k_3\cdot \hat{p}_{14}-k_3\cdot k_{12}\  p_{13}\cdot\hat{p}_{14})
\nonumber \\ & & 
\times (k_2\cdot k_{34}\ k_1\cdot k_4+k_1\cdot k_{34}\ k_2\cdot k_4-k_4\cdot k_{34}\ k_1\cdot k_2)\nonumber\\
&+&(k_{12}\cdot k_{34}\  p_{13}\cdot\hat{p}_{14}-k_{34}\cdot p_{13}\
k_{12}\cdot \hat{p}_{14}-k_{12}\cdot p_{13}\  k_{34}\cdot\hat{p}_{14})
\nonumber \\ & & \times (k_1\cdot k_2\ k_3\cdot k_4-k_1\cdot k_3\ k_2\cdot k_4-k_1\cdot k_4\ k_2\cdot k_3)\nonumber\\
&+&(k_{34}\cdot p_{13}\  k_1\cdot\hat{p}_{14}+k_1\cdot p_{13}\ k_{34}\cdot \hat{p}_{14}-k_1\cdot k_{34}\  p_{13}\cdot\hat{p}_{14})(k_4\cdot k_{12}\ k_2\cdot k_3-k_2\cdot k_{12}\ k_3\cdot k_4)\nonumber\\
&+&(k_1\cdot k_{34}\ k_3\cdot k_{12}+k_1\cdot k_3\ k_{12}\cdot k_{34}-k_1\cdot k_{12}\  k_3\cdot k_{34})(k_2\cdot \hat{p}_{14}\ k_4\cdot p_{13}+k_2\cdot p_{13}\ k_4\cdot \hat{p}_{14})\nonumber\\
&+&(k_2\cdot k_{12}\ k_4\cdot k_{34}-k_4\cdot k_{12}\ k_2\cdot
k_{34})(k_1\cdot k_3\  p_{13}\cdot\hat{p}_{14}-k_1\cdot p_{13}\
k_3\cdot \hat{p}_{14})
\nonumber \\
&-&2k_1\cdot k_4\ k_3\cdot k_{34}\ k_2\cdot p_{13}\  k_{12}\cdot\hat{p}_{14}\nonumber\\
&-&2 k_1\cdot p_{13}\ k_3\cdot k_{34}(k_2\cdot k_4\  k_{12}\cdot\hat{p}_{14}+k_4\cdot k_{12}\  k_2\cdot\hat{p}_{14}) \nonumber \\
&+& k_1\cdot \hat{p}_{14}\ k_4\cdot k_{12}(k_2\cdot k_{34}\  k_3\cdot p_{13}-2k_2\cdot p_{13}\  k_3\cdot k_{34})\nonumber\\
&+&k_2\cdot k_4\ k_{12}\cdot k_{34}(k_3\cdot p_{13}\  k_1\cdot\hat{p}_{14}+k_1\cdot p_{13}\  k_3\cdot\hat{p}_{14}) \nonumber \\
&+&2 k_1\cdot k_4\ k_{12}\cdot p_{13}(k_2\cdot k_{34}\  k_3\cdot\hat{p}_{14}-k_3\cdot k_{34}\  k_2\cdot\hat{p}_{14})\nonumber\\
&+&2k_1\cdot k_{12}\ k_4\cdot k_{34}(k_3\cdot p_{13}\
k_2\cdot\hat{p}_{14}+k_2\cdot p_{13}\  k_3\cdot\hat{p}_{14}) \nonumber \\
&+&2k_3\cdot \hat{p}_{14}\ k_{12}\cdot p_{13}(k_1\cdot k_{34}\  k_2\cdot k_4-k_4\cdot k_{34}\  k_1\cdot k_2)\nonumber\\
&+&k_1\cdot \hat{p}_{14}\ k_2\cdot k_4(k_3\cdot k_{12}\  k_{34}\cdot p_{13}-2k_3\cdot k_{34}\  k_{12}\cdot p_{13}) \nonumber \\
&+&2 k_1\cdot k_{12}\ k_3\cdot k_4(k_{34}\cdot p_{13}\  k_2\cdot\hat{p}_{14}+k_2\cdot p_{13}\  k_{34}\cdot\hat{p}_{14})\nonumber\\
&+&k_2\cdot k_4(k_{34}\cdot \hat{p}_{14}\ k_3\cdot k_{12}\ k_1\cdot p_{13}+p_{13}\cdot \hat{p}_{14}\ k_1\cdot k_{12}\ k_3\cdot k_{34})\nonumber\\
&-&k_1\cdot \hat{p}_{14}\ k_2\cdot k_{12}\ k_4\cdot k_{34}\ k_3\cdot p_{13}\,,
\end{eqnarray}
where $k_{ab}\equiv k_a-k_b$, $\hat{k}_{24}\equiv k_2+k_4$, $p_{13}\equiv p-k_1-k_3$ and $\hat{p}_{14}\equiv p-k_1+k_2-k_3+k_4$.
By means of IBPs, we express the 2-point amplitude in terms of MIs,
\begin{eqnarray}
\begin{minipage}{2.0cm}
\begin{center}
\includegraphics[width=1.8cm]{topologies/amp50}
\end{center}
\end{minipage} 
\vspace{-1cm}[N_{50}] 
&=&c_1
\begin{minipage}{2.0cm}
\begin{center}
\includegraphics[width=1.9cm]{1040802}
\end{center}
\end{minipage}
+
c_2
\begin{minipage}{2.0cm}
\begin{center}
\includegraphics[width=1.9cm]{1040603}
\end{center}
\end{minipage}
+
c_3
\begin{minipage}{2.0cm}
\begin{center}
\includegraphics[width=1.9cm]{1040607}
\end{center}
\end{minipage}
+
\nonumber \\
&+& 
c_4
\begin{minipage}{2.0cm}
\begin{center}
\includegraphics[width=1.9cm]{1040602}
\end{center}
\end{minipage}
+
c_5
\begin{minipage}{2.0cm}
\begin{center}
\includegraphics[width=1.9cm]{1040501}
\end{center}
\end{minipage}
\end{eqnarray}
and
\begin{eqnarray} 
c_1&=&-{(d-2)\*(3\*d-10)\*(3\*d^3-41\*d^2+165\*d-204)\*s^3\over4\*(d-3)\*(2\*d-7)\*(5\*d-16)\*(5\*d-14)\*(5\*d-12)},\\
c_2 &=& {(d-2)\*(51\*d^4-769\*d^3+4018\*d^2-8868\*d+7080)\*s\over2\*(d-4)^2\*(2\*d-5)\*(3\*d-10)\*(5\*d-12)},\;\;\\
c_3&=&{(d-2)\*(164\*d^5-3543\*d^4+26298\*d^3-90056\*d^2+146592\*d-92160)\*s\over12\*(d-4)^2\*(d-3)\*(5\*d-16)\*(5\*d-14)\*(5\*d-12)},\\
c_4 &=& -{(d-2)\*(9\*d-23)\*(d^2-12\*d+24)\*s\over 2(d-4)^2\*(d-3)\*(5\*d-12)},\\
c_5&=&-{(d-2)\*(609\*d^5-8946\*d^4+52176\*d^3-151096\*d^2+217360\*d-124320)\over2\*(d-4)^3\*(d-3)^2\*(3\*d-10)\*(3\*d-8)}.
\end{eqnarray} 
This result can be expanded around $d=3+\varepsilon$, using the
expressions of the MIs given in Appendix A,
\begin{eqnarray}
\mathcal{A}_{50}&=&-\text{i}
(8\*\pi\*G_N)^5\*(m_1\*m_2)^3\*2^{-4}\*(4\*\pi)^{-(4+2\*\vep)}\*e^{2\*\vep\*\gamma_E}\* s^{(1+2\*\vep)}
\times 
\nonumber\\&&
\qquad \left[
 {1\over\vep}\*\left({31\over36}-{\pi^2\over12}\right)
- {985\over216} 
+ {61\pi^2\over144} 
- {3\pi^2\over4}\*\log{2}
+ {37\zeta_3\over8}
+ \mathcal{O}(\vep)
\right].
\end{eqnarray}
Finally, by applying the Fourier transform formula (\ref{Fourier}) to $-\text{i} {\cal A}_{50}$, one gets the result for ${\cal L}_{50}$ reported in appendix B.

\bibliographystyle{JHEPmod}
\bibliography{references}

\end{document}